\documentclass{aa}
\usepackage{graphicx}
\usepackage{txfonts}
\usepackage{natbib}
\bibpunct{(}{)}{;}{a}{}{,}    %to follow the A&A style 
\newcommand{\Sect}[1]{Sect.~\ref{#1}}
\newcommand{\Fig}[1]{Fig.~\ref{#1}}
\newcommand{\Figs}[1]{Figs.~\ref{#1}}
\newcommand{\Eq}[1]{Eq.~(\ref{#1})}
\newcommand{\url}[1]{{\tt #1}}
\begin{document}
\title{The dynamical role of the circumplanetary disc in planetary migration.}

\author{
Aur\'elien Crida \inst{1,2},
Cl\'ement Baruteau \inst{3,4},
Wilhelm Kley \inst{1},
\and
Fr\'ed\'eric Masset \inst{3,5}
}
\offprints{A. Crida,\\ \texttt{crida@tat.physik.uni-tuebingen.de}}
\institute{
     Institut f\"ur Astronomie \& Astrophysik,
     Abt. Computational Physics,
     Universit\"at T\"ubingen,
     Auf der Morgenstelle 10,
     D-72076 T\"ubingen, Germany
\and
     now at\,: DAMTP, University of Cambridge,
     Wilberforce Road, Cambridge CB3 0WA, UK
\and
     Laboratoire AIM-UMR 7158, CEA/CNRS/Universit\'e Paris Diderot,
     IRFU/Service d'Astrophysique,
     CEA/Saclay,
     91191 Gif-sur-Yvette Cedex, France
\and 
     now at\,: Astronomy and Astrophysics Department, University of California,
     Santa Cruz, CA 95064, USA
\and
     ICF-UNAM, Av. Universidad s/n, Cuernavaca, Morelos, C.P. 62210, Mexico
}
\date{\today}
\abstract
%%  Context
{Numerical simulations of planets embedded in protoplanetary gaseous
discs are a precious tool for studying the planetary migration\,; however,
some approximations have to be made. Most often, the selfgravity of
the gas is neglected. In that case, it is not clear in the literature
how the material inside the Roche lobe of the planet should be taken
into account.}
%%  Aims
{Here, we want to address this issue by studying the influence of
various methods so far used by different authors on the migration
rate.}
%% Methods
{We performed high-resolution numerical simulations of giant planets
embedded in discs. We compared the migration rates with and without gas
selfgravity, testing various ways of taking the circum-planetary disc
(CPD) into account.}
%% Results
{Different methods lead to significantly different migration
rates. Adding the mass of the CPD to the perturbing mass of the planet
accelerates the migration. Excluding a part of the Hill sphere is a
very touchy parameter that may lead to an artificial suppression of
the type~III, runaway migration. In fact, the CPD is smaller than the
Hill sphere. We recommend excluding no more than a $0.6$ Hill radius
and using a smooth filter. Alternatively, the CPD can be given the
acceleration felt by the planet from the rest of the protoplanetary
disc.}
%% Conclusions
{The gas inside the Roche lobe of the planet should be very carefully
taken into account in numerical simulations without any selfgravity of
the gas. The entire Hill sphere should not be excluded. The method
used should be explicitly given. However, no method is equivalent to
computing the full selfgravity of the gas.}
\maketitle
\markboth
{Crida et al.\,: Dynamical role of the circum-planetary disc}
{Crida et al.\,: Dynamical role of the circum-planetary disc}

\section{Introduction}
\label{sec:introduction}

Planetary migration has been widely studied in the past decade, after
the discovery of the first exoplanets. Indeed, the first known
exoplanets were mostly hot Jupiters that could not have formed where
they presently orbit, according to the standard theory of planet
formation. Consequently, they are supposed to have migrated in the
protoplanetary gaseous disc before its dissipation. More recently,
pairs of exoplanets in mean motion resonance have been detected. This
particular configuration requires some dissipative process for the
planets to approach each other and lock in resonance. This is
a strong evidence for planetary migration.

The theory of planetary migration was in fact proposed before the
observational evidence
\citep{GT79,LinPapaloizou1979,MeyerVernetSicardy1987,LinPapaloizou1986b,Ward1986}. Numerical
simulations allow study of more complex cases (several planets for
instance) and refinement of our understanding of this
phenomenon. Therefore, numerical studies of planetary migration are
numerous in the literature. \citet{DDeValBorro-etal-2006} have
compared the results of various codes on two typical test problems and
find an overall consistency between the codes\,; however, a few
numerical and physical issues are still open. In particular, a
circum-planetary disc (CPD) generally appears around giant
planets. Without mesh refinement, this structure is generally poorly
resolved, but still present. As it lies close to the planet, it could
potentially exert a strong force on it. So far, a very limited number
of studies have considered the probably important role of the CPD in
the migration rate. As this material is considered to belong to the
planet and in order to avoid numerical artifacts due to the small
distance, some authors exclude the Hill sphere of the planet from the
calculation of the force of the gas on the planet. Most authors
identify the CPD with the Hill sphere, which in fact does not hold as
we shall see below \citep[see also][]{DAngelo-etal-2005}. Some authors
add the mass of the gas in the Hill sphere to the planet mass when
computing the planet gravity field. Others add the acceleration caused
by the entire gas disc on the planet to the gas in the Hill sphere. In
fact, the situation is confused, and there is no consensus on this
issue.

Recently, \citet{Crida-etal-2008a} have shown how planetary migration
can explain the orbital parameters of exoplanets in mean motion
resonance, and remarked that the outcome of their simulations had a
dependence on the way they deal with the gas in the Hill sphere of the
planets. \citet{Peplinski-etal-2008-I} have also enlightened that the
role of the CPD is not a minor one, and that a careful study of this
problem is needed to answer this question.

In this paper, we study the influence of the CPD on the planetary
migration in numerical simulations. The theory is discussed, to get a
better understanding of the problem. This leads to possible solutions
to treat the CPD. We perform 2D simulations with modest to high
resolution, and we test and compare various recipes. Our goal is to
provide a survey of the parameter ``CPD'' in numerical simulations. In
particular, the fraction of the Hill sphere that is excluded is a very
touchy parameter, and the way the exclusion is performed has a great
influence on the outcome of the simulation. Many authors are not aware
of this. Our survey shows that some popular methods are
inappropriate. We present better ways of dealing with the material
present in the Hill sphere, in particular the CPD.

This requires an analysis of the Hill sphere structure, but in the
framework of global disc simulations, which are not dedicated to
resolve the Hill sphere with a high level of precision. We do not aim
at understanding the detailed physics and the fine structure of the
CPD itself, but more how the gas in the Hill sphere of the planet
should be taken into account. Convergence properties and the impact of
varied prescriptions on the equation of state (EOS) or the softening
length are fundamental issues, but we do not address them in this
work. Instead, our aim is to find an adequate prescription for the
torque calculation in non selfgravitating simulations, based on
general considerations such as the conservation of angular momentum
and the main features of the flow topology in the vicinity of the
planet. The conservation of angular momentum is enforced in our scheme
independently of the resolution, while the flow topology depends very
weakly on the resolution (even if the mass of the CPD does, in an
EOS-dependent way).

This paper is organised as follows. In \Sect{sec:pb}, the problem is
explained and analysed. Some definitions are provided, in order to
clarify the strategy and the questions that arise from the presence of
a CPD. We show that the problem can be divided into the questions of
the {\it perturbing mass}, of the {\it gravitational mass}, and of the
{\it inertial mass} of the planet. In \Sect{sec:code}, the numerical
setup is presented. Then, in the following sections, numerical
experiments are performed. In \Sect{sec:numII}, a Jupiter mass planet,
in type~II migration in considered. In \Sect{sec:numIII}, the
influence of the CPD on a Saturn mass planet, in type~III migration is
studied. In \Sect{sec:direct}, the structure of the Hill sphere is
analysed in detail, and the size of the CPD is estimated. Last, our
results are summarised, and we conclude in \Sect{sec:conclu}.

\section{Problem presentation}
\label{sec:pb}

\subsection{Disc structure}
\label{subsect:structure}

Let us consider a planet on a circular orbit around a central star. In
steady state, in the frame corotating and drifting along with
the planet, the gaseous disc is split into 4 closed regions\,:
\begin{enumerate}
\item The {\it Roche lobe} of the planet, which is where the
streamlines are closed around the planet. In the restricted three-body
problem, it can be approximated by the {\it Hill sphere}, centred on
the planet, of radius the Hill radius\,: $r_H = a_p (q/3)^{1/3}$,
which is where $a_p$ is the semi-major axis of the planet and
$q=M_p/M_*$ is the ratio of the planet and stellar masses.
\item The {\it horseshoe region} around the orbit of the planet,
where the streamlines have a horseshoe shape. This region is delimited
by the horseshoe {\it separatrices}.
\item The {\it inner disc} between the star and the horseshoe inner separatrix.
\item The {\it outer disc} beyond the horseshoe outer separatrix.
\end{enumerate}
The Roche Lobe disappears for low-mass planets \citep[$q
\lesssim 10^{-4}$, but this depends on the aspect ratio,
see][Fig.~11]{Masset-etal-2006b}.

The gas in the inner disc and in the outer disc exert a
torque on the planet, the {\it differential Lindblad torque}. The gas present in the
horseshoe region exerts the so-called {\it horseshoe drag}
\citep{Ward1991,Masset2001}. For low-mass planets in locally
isothermal discs, the sum of these two effects is generally a negative
torque and responsible for type~I planetary migration
\citep{Ward1997}. If the energy equation is taken into account, the
torque due to the horseshoe drag is modified in radiatively
inefficient discs as well as in more realistic radiative discs. This
may slow down or reverse the type~I migration
\citep{PaardekooperMellema2006,BaruteauMasset2008AD,KleyCrida2008}. For
intermediate mass planets, under some circumstances, the horseshoe
drag can also be responsible for type~III migration
\citep{MassetPapaloizou2003}.

The effect of the gas present in the Roche lobe of the planet has so
far been neglected by most authors \citep[with the noticeable
exceptions of][see below]{DAngelo-etal-2005,Peplinski-etal-2008-I},
and never explicitly analysed. However, as this material is located
very close to the planet, it could exert on it a strong force and
torque. Consequently, it has to be carefully taken into account. We
show further that it can influence the migration rate by almost a
factor $2$. This concerns planets of sufficient mass, which are able to
create a Roche lobe in the disc and to accrete gas.

Please note that the CPD does not necessarily fill the entire Hill
sphere. There may be material inside the Hill sphere that is not
orbiting around the planet, but that is simply passing
by. \citet{DAngelo-etal-2005,Peplinski-etal-2008-I} have shown that
the gas dynamics around a migrating planet is more complicated than
described in \Sect{subsect:structure}, and that the Roche lobe may be
modified. This may be a source of confusion. Hereafter, the CPD always
refers to the material bound to the planet --\,that is\,: on a
streamline that describes an orbit around the planet\,-- while the
Hill sphere refers to the material located within a distance $r_H$ to
the planet.

\subsection{Gravitational interactions}

\begin{centering}
\begin{figure}
\includegraphics[width=\linewidth]{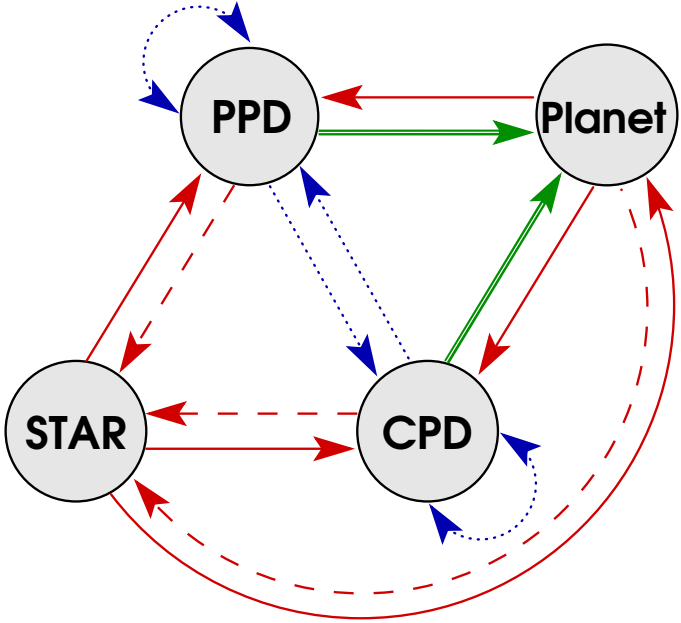}
\caption{Summary of the gravitational interactions in the problem of a
planet, surrounded by a CPD, orbiting in a PPD, about a star.}
\label{fig:schema}
\end{figure}
\end{centering}

In \Fig{fig:schema} are displayed all the gravitational interactions
physically present in the problem. The gaseous disc has been split
into the CPD, which is the gas orbiting the
planet, bound to the planet, and the rest of the ProtoPlanetary Disc
(PPD). Four systems are interacting with each other, which makes a
total of 12 interactions. In addition, the PPD and the CPD, as
extended systems, exert on themselves a gravity force.

These 14 gravitational actions can be classified in 4 categories\,:
\begin{itemize}
\item {\bf 5 direct terms} (red solid arrows)\,: The Planet$\to$PPD,
Planet$\to$CPD, Star$\to$PPD, Star$\to$CPD, Star$\to$Planet are the
most essential interactions of the system. They are responsible for
the disc and planet orbiting the star and for the perturbations of the
disc, including the mere existence of the CPD.
\item {\bf 3 indirect terms} (red dashed arrows)\,: The (re)actions of
the PPD, the CPD, and the planet on the star are generally considered
as indirect terms in the expression of the potential, if the frame is
centred on the star.
\item {\bf 4 selfgravity terms} (blue dotted arrows)\,: All the
interactions between parts of the gaseous disc belong to what is
called the selfgravity of the disc (PPD$\to$PPD, PPD$\to$CPD,
CPD$\to$PPD, CPD$\to$CPD).
\item {\bf 2 migration terms} (green double arrows)\,: The
action of the disc (PPD and CPD) on the planet are responsible for the
planetary migration.
\end{itemize}

If all the 14 interactions were taken into account in the simulations,
the computations would be self-consistent. The interactions of the two
first categories are always taken into account and are clearly
understood. But the selfgravity is generally not computed, because it
is too expensive in terms of CPU-time. Consequently, the 2 migration
terms may be inaccurate.

The PPD$\longleftrightarrow$PPD effects, which is the influence of the
gas selfgravity on the differential Lindblad torque, has been studied
for low-mass planets by \citet{NelsonBenz2003I},
\citet{PierensHure2005}, and more recently by
\citet{BaruteauMasset2008DDA,BaruteauMasset2008SG}. They confirmed
that the type~I migration is slightly accelerated by the disc
selfgravity. This effect is not discussed here. On the contrary, we
only focus on the effect of the CPD.

In this paper, we plan to analyse the role and the behaviour of the
CPD in the migration, which is represented by the arrows
CPD$\to$Planet, CPD$\to$PPD, and PPD$\to$CPD. As already said, the
action Planet$\to$CPD should always be present and is not in
question. In simulations with full gas selfgravity computed, all the
blue dotted arrows of \Fig{fig:schema} are taken into account,
including the CPD$\leftrightarrow$CPD arrow. However, the selfgravity
of the CPD should only give a minor change of the CPD structure. It
should therefore be negligible for what concerns the planetary
migration. In what follows, we do not discuss the role of the
CPD$\leftrightarrow$CPD gravitational interaction.

\subsection{Effect of the circumplanetary disc on the planet}
\label{subsec:effects}

The presence of a CPD, gravitationally bound to the planet, can affect
the migration by three means\,:
\begin{enumerate}
\item CPD$\to$Planet\,: External perturbations (Star$\to$CPD,
PPD$\to$CPD) make the CPD not axisymmetric around the planet. Thus,
the CPD could exert a torque on the planet\,; we call this torque due
to the CPD structure a {\it direct torque}.
\item CPD$\to$PPD$\to$Planet\,: By perturbing the PPD, the CPD
influences the action of the latter on the planet. Basically, the CPD
acts like the planet does, enhancing the amplitude of the
perturbation. We define this as the {\it perturbing mass problem}.
\item CPD$\to$Planet in migration\,: The planet and all the material
that is bound to it migrate together. During the migration, the CPD is
still linked to the planet. Consequently, the CPD is like a ball and
chain for the planet. If the full gas selfgravity is computed, the
CPD feels almost the same specific torque from the PPD than the
planet\,; then, the ball is pulled by the PPD, together with the
planet, and the chain is not taut. If the full gas selfgravity is not
computed (no blue arrow in \Fig{fig:schema}), the CPD feels no torque
from the PPD and tends to stay on a constant orbit. Then, the
migrating planet has to pull the CPD. The chain is taut, which slows
down the planetary migration. We define this as the {\it inertial mass
problem}.
\end{enumerate}

The first and third cases are included in the green, double arrow
CPD$\to$Planet in \Fig{fig:schema}. The force acting on the planet in
these two cases arises from gravitational interaction with the gas
located in the CPD itself.

The second case is an indirect interaction, represented in
\Fig{fig:schema} by the two-arrow path\,: CPD$\to$PPD$\to$Planet. The
action of the CPD on the PPD (blue dotted arrow CPD$\to$PPD) may change
the value of the green double arrow PPD$\to$Planet.

\subsection{Problem analysis and possible solutions}
\label{subsec:specific}

In this paper, we evaluate either the specific torque felt by the
planet $T_p$ or the migration rate $\dot{a}_p$ (the dot denotes the
derivative with respect to time). Both quantities are equivalent, as
for a planet on a circular orbit, $2a_p\dot{a}_pB=T_p$, with $B$ the
second Oort constant.

It is well-known that a planet orbiting in a gaseous PPD perturbs the
latter gravitationally. In the linear regime, this perturbation leads
to the appearance of a one-armed spiral wave, called the \emph{wake},
in the inner and the outer discs. The amplitude of the perturbation is
proportional to the planet mass $M_p$. The perturbed density
distribution leads to a gravitational force on the planet. This force
is proportional to the mass of the wake times the mass of the planet
{\it i.e.}\ proportional to ${M_p}^2$. Thus, the planet feels from the
disc a torque proportional to its mass squared. As a result, the
specific torque applied to the planet is proportional to the planet
mass $M_p$. This corresponds to the PPD$\leftrightarrows$Planet arrows
in Fig.~\ref{fig:schema}.

If the planet is surrounded by a CPD, things are
slightly different. Indeed, the CPD is gravitationally bound to the
planet, and it moves with the planet. Thus, the planet of the above
paragraph should be replaced by the planet and its CPD. The specific
torque felt by this system is proportional to $(M_p+M_{\rm CPD})$,
where $M_{\rm CPD}$ is the mass of the CPD. The presence of the CPD
should increase the migration speed.

More generally, the specific torque felt by the migrating body is
proportional to the amplitude of the perturbation of the disc
(proportional to the {\it perturbing mass} of the planet), times the
mass of the body gravitationally interacting with it ({\it
gravitational mass} of the planet), divided by the mass of the
migrating body ({\it inertial mass} of the planet). If selfgravity is
computed, these three masses are equal to $(M_p+M_{\rm
CPD})$. However, as we see below, the perturbing, gravitational,
and inertial masses of the planet differ if the gas selfgravity is
not computed. This is the problem.

In numerical simulations without gas selfgravity, the CPD does not
perturb the PPD, neither does it feel a torque from
it\,: in \Fig{fig:schema}, the blue dotted arrows are suppressed. The
perturbing mass is $M_p$, and the force felt by the planet is
proportional to ${M_p}^2$. But the CPD is linked to the planet. To
follow the planet in its migration, the CPD has to exchange angular
momentum with it. Thus, the relevant inertial mass of the planet is
not $M_p$ but $(M_p+M_{\rm CPD})$. The specific torque felt by the
migrating system (the planet and its CPD) is proportional to
${M_p}^2/(M_p+M_{\rm CPD})$. Here, the presence of the CPD decreases
the migration speed, which should not occur.

Based on this analysis, several authors
\citep[e.g.][]{PierensNelson2008} have tried to cancel the force from
the CPD on the planet by excluding the Hill sphere of the computation
of the force of the disc on the planet. This aims at suppressing the
green double arrow CPD$\to$Planet in \Fig{fig:schema}. Thus, the
inertial mass of the planet is $M_p$ again, and one is back to the
first case\,; the specific torque is proportional to $M_p$, but not to
$(M_p+M_{\rm CPD})$.

Others \citep[e.g.][]{Masset2006Oleron,Peplinski-etal-2008-I} have
suggested adding $M_{\rm CPD}$ to the perturbing mass of the planet
(denoted $M_p^*$ by \citeauthor{Peplinski-etal-2008-I}), to make the
amplitude of the wake proportional to $(M_p+M_{\rm CPD})$. In
\Fig{fig:schema}, this is equivalent to adding the blue, dotted
CPD$\to$PPD arrow. This gives a torque proportional to $(M_p+M_{\rm
CPD})M_p$ and a specific torque proportional to $(M_p+M_{\rm
CPD})M_p/(M_p+M_{\rm CPD})=M_p$ as well. \citet{Peplinski-etal-2008-I}
find very little change in the migration rate by using $M_p$ or
$(M_p+M_{\rm CPD})$ as the perturbing mass of the planet, which is
quite surprising, as they have $M_{\rm CPD}\sim M_p$.

By adding $M_{\rm CPD}$ to the planet mass felt by the disc (the
perturbing mass) and excluding the CPD from the disc felt by the
planet, one should recover a specific torque proportional to
$(M_p+M_{\rm CPD})M_p/M_p=(M_p+M_{\rm CPD})$. One could also think of
adding $M_{\rm CPD}$ to the perturbing mass of the planet and to the
gravitational mass of the planet. In that case, the torque is
proportional to $(M_p+M_{\rm CPD})^2$. As the inertial mass is
$(M_p+M_{\rm CPD})$, we find a specific torque in $(M_p+M_{\rm CPD})$,
as required.

\citet{Peplinski-etal-2008-I} have already addressed this issue and
quite clearly described the problem. In the frame of type~III
migration, they argue against any exclusion of a part of the disc in
the torque calculation, but they propose another solution by applying
the acceleration felt by the planet from the disc to the CPD (see
their Eqs.~(13) and (17)\,). This should mimic the blue, dotted arrow
PPD$\to$CPD. In their paper, this acceleration is given by all the
disc, CPD included. Then, if the CPD exerts, say, a positive torque on
the planet for whatever reason, this positive contribution will be
applied to the CPD as well. This should not be, because it opens the
possibility for the CPD to pull the planet indefinitely, without
losing any angular momentum. For consistency, the acceleration applied
to the CPD should be the one felt by the planet from the PPD, CPD
excluded\,: the green, double arrow PPD$\to$Planet in our
\Fig{fig:schema}.

The giant planets studied in this paper are not in the linear regime,
so that the torque is not exactly proportional to ${M_p}^2$. However,
this approximation allows the above reasoning. It enlightens the role
of the CPD in the migration process and suggests solutions. In
Sects.~\ref{sec:numII} and \ref{sec:numIII}, we numerically test these
recipes and compare them with the results obtained with full
selfgravity.

\section{Code and setting description}
\label{sec:code}

To compare the migration rates obtained with different
prescriptions to take the CPD into account, we performed numerical
simulations, using the code FARGO \citep{FARGO,FARGO2}. This is a
publicly available\footnote{\tt http://fargo.in2p3.fr/} 2D code,
using polar coordinates $(r,\theta)$. We used the ADSG version, in
which the energy equation and the gas selfgravity are
implemented. The gas selfgravity is described in
\citet{BaruteauMasset2008SG}. It can be turned on or off.

The uniform kinematic viscosity is $\nu = 10^{-5}\ {a_p}^2\Omega_p$
($\Omega_p$ being the angular velocity of the planet and $a_p$
its semi major axis). The initial aspect ratio is $h_0 = (H/r)_0 =
0.03$. To take the disc thickness into account, a smoothing length
$\epsilon=0.018\,a_p$ ($60\%$ of the initial disc height) is used in
the expression of the planet potential $\Phi_p$\,:
\begin{equation}
\Phi_p = - \frac{GM_p^*}{s'}\ ,
\label{eq:Phi_p}
\end{equation}
where $s'=\sqrt{s^2+\epsilon^2}$ with $s$ the distance to the planet,
and $M_p^*$ is the perturbing mass of the planet, equal to $M_p$ if
not otherwise specified.

\begin{centering}
\begin{figure}
\includegraphics[angle=270,width=\linewidth]{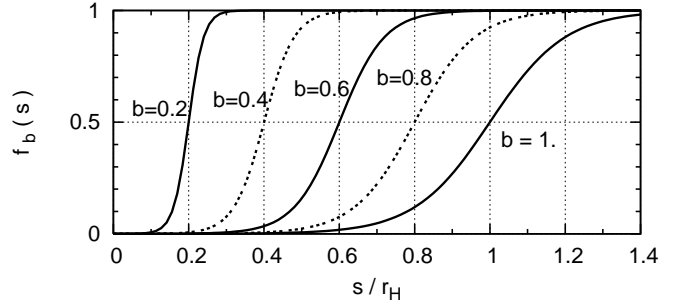}
\caption{The function $f_b(s)$ given by \Eq{eq:f} for various values of $b$.}
\label{fig:f_b}
\end{figure}
\end{centering}

The vector force $\vec{F_p}_{,\,b}$ exerted by the disc on the planet
is computed using the following expression\,:
\begin{equation}
\vec{F_p}_{,\,b} = \int_{s=0}^{+\infty}\int_{\theta=0}^{2\pi}
\frac{GM_p\Sigma}{s'^3}f_b(s) \vec{s}\,{\rm d}\theta\,{\rm d}s\ ,
\label{eq:Force}
\end{equation}
where $\vec{s}$ is the vector originating from the planet to the
centre of the considered cell. The force $\vec{F_p}_{,\,b}$ depends of
the parameter $b$ through the term $f_b(s)$\,: this is our filter used
to exclude smoothly the neighbourhood of the planet, if necessary,
also used in \citet{Crida-etal-2008a}. It is a smooth increasing
function from 0 at $s=0$ to $1$ when $s\to\infty$ through $1/2$ when
$s=b\,r_H$, drawn in \Fig{fig:f_b}, and given by:
\begin{equation}
f_b(s)=\left[\exp\left(-10\left(\frac{s}{br_H}-1\right)\right)+1\right]^{-1}\ .
\label{eq:f}
\end{equation}

For $s<br_H$, $f_b(s)\ll 1$, while $f_b(s)\sim 1$ for $s>br_H$, so
that the gas within a distance $br_H$ of the planet is neglected while
the gas outside is taken into account. For $b=0$, $f_0\equiv 1$, so
that all the disc is taken into account\,; $b=0$ gives the standard
expression of the force, with no exclusion of any part of the disc.

The planet is not accreting.

In the energy equation, the heat source is the viscous dissipation,
and the cooling follows the following law, different from the public
version of FARGO-ADSG\,:
\begin{equation}
\frac{\partial e}{\partial t} = -\sigma_R T^4/\kappa\Sigma\ ,
\label{eq:cool}
\end{equation}
where $e$ is the thermal energy of the gas per unit area, $\sigma_R$
is the Stefan-Boltzmann constant, $T$ is the temperature, $\Sigma$ is
the surface density of the gas, and $\kappa$ is the opacity, given
by\,:
\begin{equation}
\kappa=\sigma_R\frac{8h_0^8}{9{\Sigma_0}^2\nu}\ .
\end{equation}
This arbitrary constant opacity $\kappa$ ensures that, in the absence
of perturbation, the initial state is in thermodynamic equilibrium
with a density profile $\Sigma(r)=\Sigma_0r^{-1/2}$. Our equation of
state is then, with $c_s$ the sound speed, $\mu$ the mean molecular
weight, $R$ the universal gas constant, and $\gamma=1.4$ the adiabatic
index\,:
\begin{equation}
P = \Sigma {c_s}^2 / \gamma = \Sigma \frac{R\,T}{\mu}\, .
\label{eq:EOS}
\end{equation}

The computation of the energy equation provides more realistic
discs. In addition, it reduces the problem of the accretion of the
CPD. Indeed, with a locally isothermal EOS, no pressure effect stops
the collapse of the CPD. Then, at high resolution, gas in the
neighbourhood of the planet reaches extremely high densities, and
convergence is hard to obtain. Since we plan to increase gradually
the resolution of our simulations (see next paragraph), we need a
viscous and a compressional heating and a radiative cooling for the
CPD structure to remain within reasonable limits of density and
temperature. In addition, preliminary tests show that with the gas's
full selfgravity and locally isothermal EOS, we could not reach
convergence of the mass inside the Hill sphere. This motivated us to
treat the gas thermodynamics. Our opacity is admittedly not realistic,
and another prescriptions for $\kappa$ would make the CPD larger or
smaller. Once again, we are not interested in the true structure of
the CPD in this study, but we simply want a stationary CPD, of
reasonable size, to check how it influences the migration.

The radial grid extent is $r\in [0.4 a_p ; 2.5a_p]$. The regions
$[0.4a_p;0.5a_p]$ and $[2.1a_p;2.5a_p]$ are wave-damping regions, in
which the density, the temperature, and the velocities are damped
toward the initial values. The rings are geometrically spaced and the
cells squared\,: $\delta r/r = \delta \theta$ is constant.  To
save computing time, the resolution is initially low, and regularly
doubled (see Table~\ref{tab:resol}). In the end, a resolution of
$N_{\rm rings} \times N_{\rm sectors}=1840\times 6288$ and $\delta
r/r=\delta\theta=10^{-3}$ is reached. To our knowledge, this is the
first time that the FARGO code is run at such a resolution. Most of
the very-high-resolution simulations were run on a parallel cluster
with 64 CPUs. The FARGO code uses a staggered mesh, with the density
and temperature evaluated at the centre of the cells, and the
velocities at the boundaries. When the resolution is doubled, each
cell is divided into four. The new values of the fields were computed
with a 2D linear interpolation on the old grid.

\begin{table}
\caption{Resolution of the simulations.}
\begin{tabular}{rlrc}
 & \multicolumn{2}{|c|}{resolution} & \\
time $[$orbits$]$ & \multicolumn{1}{|l}{$N_{\rm rings} \times N_{\rm sectors}$}
& $\delta r/r=\delta\theta$ & \multicolumn{1}{|c}{designation} \\
\hline
\hline
from   0 to 150 : & 115 x 393 & $0.016$ & initialisation\\
from 150 to 250 : & 230 x 786 & $8 \times 10^{-3}$ & low\\
from 250 to 325 : & 460 x 1572 & $4 \times 10^{-3}$ & middle\\
from 325 to 400 : & 920 x 3144 & $2\times 10^{-3}$ & high\\
from 400 to 477 : & 1840 x 6288 & $10^{-3}$ & very high\\
\end{tabular}
\label{tab:resol}
\end{table}

Some remarks concerning the gas selfgravity are in order. The ADSG
version of the FARGO hydro-code enables switching on or off the gas
selfgravity, the computation of which is described in
\citet{BaruteauMasset2008SG}. In addition, it is possible to compute
only the axisymmetric component of the gas gravity. This is the
gravity field caused by the azimuthally averaged density field of the
gas, or by an axisymmetric disc with the same density profile as the
actual gas disc. This component of the gas gravity field is
responsible for increasing the angular velocity of the planet and of
the gas itself (at least far enough from the disc inner edge). If one
abruptly switches on the gas selfgravity when restarting a simulation
computed without selfgravity, a discontinuity in the velocities is
introduced. To avoid this, all our simulations are computed with the
axisymmetric component of the gas gravity taken into account. This has
a negligible computational cost, while the full selfgravity is quite
expensive. As discussed in \citet{BaruteauMasset2008SG}, accounting
for the axisymmetric component of the disc gravity has only a marginal
effect on the Lindblad torque with respect to the situation where the
planet is held on a fixed circular orbit in a non selfgravitating
disc. In the cases that we study here, the planet is massive enough to
shape a (deep or shallow) gap in the disc and to have a
CPD. Consequently, the effect of selfgravity is different from in the
above-mentioned studies. The shift of Lindblad resonances --\,which is
essentially the only effect in type I migration\,-- is taken over by
the question of the gap shape, which gives the gas density at the
position of each resonance, hence the torque. In addition, there are
other issues, such as a modification of the perturbing, gravitational,
and inertial mass of the planet, which we will address here.

\section{Type~II migration case}
\label{sec:numII}

Following the procedure described in \Sect{sec:code}, we performed
simulations of a Jupiter mass planet, with gas initial surface
density\,: $\Sigma_0(r)=10^{-4} (r/a_p)^{-1/2}\ M_*/{a_p}^2$. In this
case, the planet opens a deep gap \citep{LinPapaloizou1986a,
Crida-etal-2006}, and the disc mass is low enough for type~III
migration not to happen \citep[Fig.~12
of][]{MassetPapaloizou2003}. Thus, the planet should migrate in
type~II migration \citep{LinPapaloizou1986b}. This choice could look
inappropriate, because in standard type~II migration, the migration
speed is independent of the planet mass. However, this stands only if
the disc is massive enough to push the planet at the viscous accretion
speed. This occurs when $\mu>q/4$, where $\mu = 4\pi {a_p}^2 \Sigma_0
/ M_*$ and $q=M_p/M_*$ \citep{Crida-Morby-2007,Crida-these}. Here,
$q>4\mu$, the disc is not massive enough for the migration to occur at
a viscous rate. Consequently the migration rate should decrease with
the inertial mass of the planet. This case is timely for studying our
problem.

\subsection{Inertial mass problem}

As explained in \Sect{subsec:effects}, this problem only concerns
migrating planets. To study this effect, we let the planet evolve
freely in the disc. To test and quantify the influence of the CPD on
the migration speed, a region around the planet is excluded from the
computation of the force of the disc on the planet. To perform this,
the parameter $b$ in \Eq{eq:Force} is taken strictly positive, and
several values are tested.

\subsubsection{Middle resolution}
\begin{centering}
\begin{figure}
\includegraphics[angle=270,width=\linewidth]{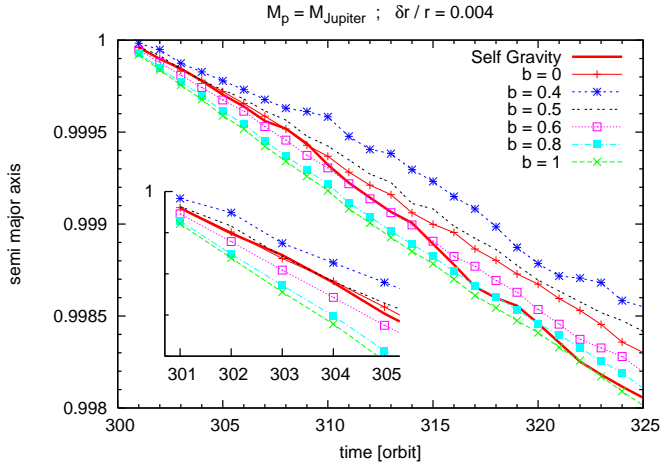}
\caption{Migration path of the Jupiter mass planet after its release
at $t=300$ orbits, with $\delta r/r = \delta \theta = 0.004$. The bold
curve corresponds to a simulation where the full gas selfgravity is
computed. The other ones come from simulations with exclusion of a
part of the Hill Sphere, for various values of $b$. The first five
orbits are enhanced in the inset at the bottom left.}
\label{fig:migr-s12_Jup}
\end{figure}
\end{centering}

First, the planet is released after 300 orbits on a fixed circular
orbit\,; the resolution of the grid is $\delta r = 4\times 10^{-3}\,r
= r_H/17.3$ (see Table~\ref{tab:resol}), which is finer than the
typical kind of resolution used in the literature. The migration is
followed for 25 orbits. Various values of $b$ from 0 to 1 are
used. The results are displayed in \Fig{fig:migr-s12_Jup}. The curves
for $b\leqslant 0.3$ almost overlap\,; therefore, only the curve
for $b = 0$ is drawn as a thin red solid line with $+$ symbols. For
$b=0.4$, the planet follows the blue starred line. It migrates
significantly more slowly, and in 25 orbits, its semi-major axis has
decreased $8.5\%$ less. For $b\geqslant 0.4$, the migration speed
increases with $b$, and for $b=1$ (exclusion of all the Hill sphere),
the semi-major axis of the planet decreases in 25 orbits nearly $12\%$
more than in the case $b=0$. The migration ratio after 10 orbits
between $b=0.4$ and $b=1$ is almost 2. This shows that the Hill sphere
can have a strong influence on the migration rate in numerical
simulations.

The bold solid red curve in \Fig{fig:migr-s12_Jup} is obtained with a
simulation where the full selfgravity of the gas has been taken into
account, while only the axisymmetric component is used in the
others. Of course, in that case, no exclusion of any part of the disc
is done, $b=0$. This bold curve should be seen as the reference.

\begin{centering}
\begin{figure}
\includegraphics[angle=270,width=\linewidth]{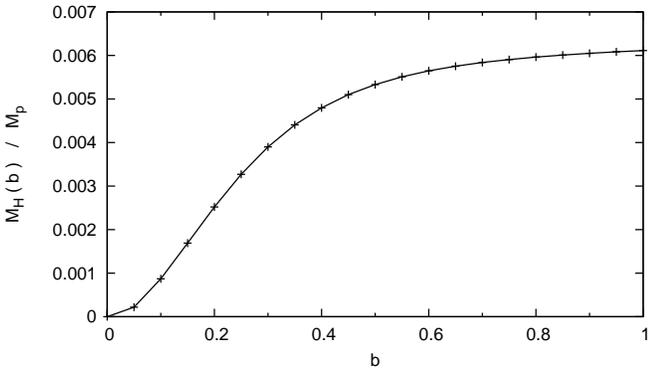}
\caption{Mass distribution inside the Hill sphere of the Jupiter mass
planet given by \Eq{eq:mass}, in planet mass units. The measure is
made after 300 orbits on a fixed circular orbit, with a resolution of
$\delta r/r = 0.004$.}
\label{fig:Mcpd-s12_Jup}
\end{figure}
\end{centering}

To better understand what happens in the Hill sphere,
\Fig{fig:Mcpd-s12_Jup} displays the mass distribution in the Hill
sphere of the planet, at the moment where the planet is released,
computed using the following expression\,:
\begin{equation}
M_{\rm H}(b) =
\int_{s=0}^{+\infty}\int_{\theta=0}^{2\pi} \Sigma\ \left(1-f_b(s)\right)
\ s{\rm d}\theta\,{\rm d}s\ .
\label{eq:mass}
\end{equation}
If, instead of the smooth $f_b$ function, a Heaviside
function was used, $M_{\rm H}(b)$ would have the more intuitive
following expression\,:
$$
M_{\rm H}(b) 
\approx \int_{s=0}^{br_H}\int_{\theta=0}^{2\pi} \Sigma 
\ s{\rm d}\theta\,{\rm d}s\ .
$$
If $\Sigma$ was constant, $M_{\rm H}(b)$ should increase as
$b^2$. This is approximately the case only for $b\leqslant 0.15$, which
shows that the density decreases strongly with $s$ for $s>0.15r_H$. In
fact, $90\%$ of the mass is inside $0.55r_H$, which favours the idea
that the CPD should be considered as only the inner half of the Hill
sphere.

The total mass in the Hill sphere is $0.61\%$ of the planet mass, so
that the inertial effect should be very small. In addition, the
variation of $M_{\rm H}(b)$ for $b>0.5$ is only a few percent, so
the $20\%$ difference in the migration rates between the cases
$b=0.5$, $0.6$, $0.8$, and $1.0$ cannot be explained by a variation in
the mass of the excluded region.

\subsubsection{High resolution}
\label{sss:HR}
\begin{centering}
\begin{figure}
\includegraphics[angle=270,width=\linewidth]{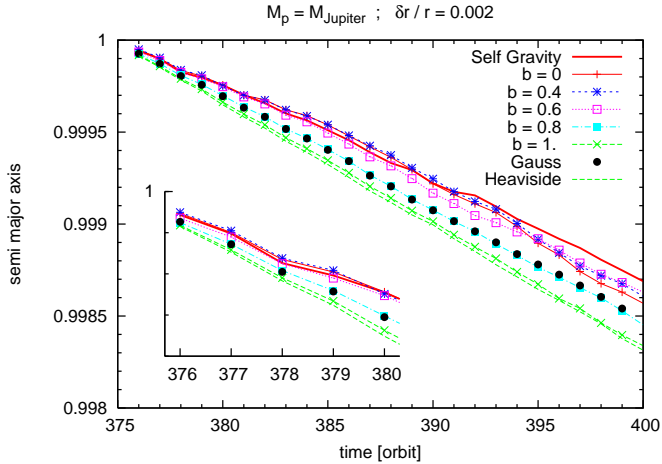}
\caption{Migration path of the Jupiter-mass planet after its release
at $t=375$ orbits, with $\delta r/r = \delta \theta = 0.002$. Same
colour code as \Fig{fig:migr-s12_Jup}. The five first orbits are
enhanced in the inset at the bottom left.}
\label{fig:migr-s15_Jup}
\end{figure}
\end{centering}

In a second experiment, the planet is not released at $t=300$, but
kept on a fixed circular orbit until $375$ orbits, the resolution
being doubled at time $325$ (see Table~\ref{tab:resol}). The planet is
then released at $375$ orbits, with a resolution of $0.002$ (35
cells in a Hill radius, $1000$ cells in the surface of a circle of
radius $0.51r_H$). The results are displayed in
\Fig{fig:migr-s15_Jup}. The curves with symbols were performed using
$f_b$ as filter, for various values of $b$.

The average migration rate is a bit slower than in the previous
case. The green dashed curves with $\times$ symbols (case $b=1$) in
\Figs{fig:migr-s12_Jup} and \ref{fig:migr-s15_Jup} are rather linear,
providing a constant migration rate, only given by the torque from the
PPD outside of the Hill sphere. This torque should be converged for
resolutions of the order of $r_H/10$ (here\,: $0.0046$), as widely
assumed in the literature \citep[e.g.][]{DAngelo-etal-2005}. However,
the gap profile is slowly evolving with time, so that the torque
exerted on the planet at $t=300$ or $375$ orbits differs,
independently on the resolution. This time delay is the reason why the
green dashed curves with $\times$ symbols slightly differ in
\Figs{fig:migr-s12_Jup} and \ref{fig:migr-s15_Jup}.

For the other curves, with narrower filters $f_b$ or full gas
selfgravity, their various evolutions between the medium-resolution
and high-resolution cases suggest that convergence in the Hill sphere
had not been reached at medium resolution. In particular, the
different variations of the cases $b=0.4$ and $b=0.6$ between
\Figs{fig:migr-s12_Jup} and \ref{fig:migr-s15_Jup} can hardly be
explained by a simple smooth time variation of the disc
structure. However, we recall that a convergence study of the CPD
structure is certainly interesting but is not within the scope of this
paper, focused as it is on the role of the CPD in migration in
standard numerical simulations.

With $\delta r/r = 0.002$, all the curves are almost linear, denoting
that the CPD structure is better resolved. Then, the role of the
filter function can be analysed better. Here, the trend is clear. The
larger $b$, the lower the inertial mass of the planet, and the
faster the migration, as expected if excluding a part of the Hill
sphere only had an effect on the inertial mass of the planet.

In addition, other shapes of filter are tested. The black dots
correspond to a case where $f_b(s)$ has been replaced in \Eq{eq:Force}
by a Gaussian filter\,:
\begin{equation}
g(s)=1-\exp(-(s/r_H)^2)\ .
\label{eq:g}
\end{equation}
It gives almost the same result than $f_{0.8}$ (light blue dot-dashed
curve with full squares). The green dashed line without $\times$
symbols corresponds to a Heaviside filter $h(s)=0$ if $s<r_H$,
$h(s)=1$ if $s>r_H$. It gives almost the same result as $f_1$ (green
dashed line with $\times$ symbols).

For small $b$, the difference between $b=0.6$, $b=0.4$, $b=0$, and the
full selfgravity case is rather small, although no match is
observed. The effect of gas selfgravity on type~II migration is little
and has nothing to do with its effect on type~I migration, which could
be expected as the torque felt by the planet in this case is equal to
the viscous torque from the disc and not the differential Lindblad
torque (which is affected by the gas full selfgravity). Higher values
of $b$, or the use of $g(s)$ or $h(s)$ show a significant acceleration
of the migration with respect to the full selfgravity case.

\begin{centering}
\begin{figure}
\includegraphics[angle=270,width=\linewidth]{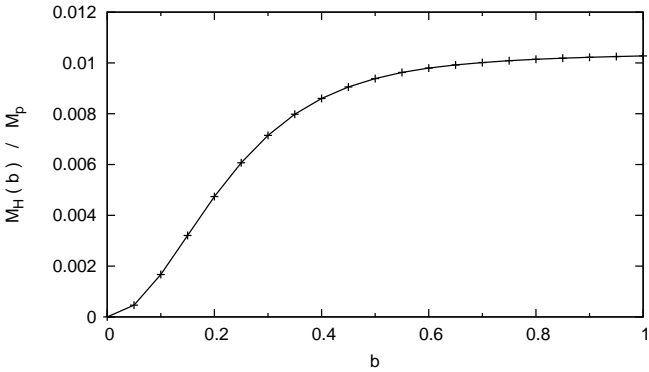}
\caption{Mass distribution inside the Hill sphere of the Jupiter mass
planet given by \Eq{eq:mass}, after 375 orbits, with a resolution of
$\delta r/r = 0.002$.}
\label{fig:Mcpd-s15_Jup}
\end{figure}
\end{centering}

Figure~\ref{fig:Mcpd-s15_Jup} shows the mass distribution inside the
Hill sphere of the planet at the moment it is released, after 375
orbits. The distribution is a bit more compact than in previous case\,:
$90\%$ of the mass is now included in $0.48r_H$. The total mass is
higher ($0.01\,M_p$). This is due to the doubling of the resolution.

The value of $M_{\rm H}(b)$ does not vary significantly for
$b>0.5$. Consequently, the inertial mass of the planet should not
change much whether one half of the Hill sphere is excluded or the
full Hill sphere. The acceleration of the migration for $b>0.6$ is
probably not due to the decrease in the inertial mass of the
planet. As in the middle resolution case, we find that by excluding
all the Hill sphere, one misses a part of the torque that this is not
an artifact from the absence of selfgravity.

\subsection{Perturbing and gravitational mass problem}
\label{subsec:grav_Jup}

The addition of the CPD mass to the perturbing mass of the planet can
be easily implemented in the code. One should simply add the mass of
the CPD to the planet perturbing mass in computating the
gravity potential of the planet through \Eq{eq:Phi_p}. However,
it is not easy to measure the mass of the CPD $M_{\rm CPD}$, because
this requires analysing the disc structure. We assume that the
CPD is a disc centred on the planet of radius
$r\approx0.6r_H$. Therefore, we take $M_{\rm CPD}=M_{\rm
H}(b=0.6)$. Figures~\ref{fig:Mcpd-s12_Jup} and \ref{fig:Mcpd-s15_Jup}
have shown that the obtained value of $M_{\rm CPD}$ does not vary by
more than $10\%$ for $0.5<b<1$. The CPD itself should not
feel its own mass as added to the planet. This would be
inconsistent. Therefore, the expression of the gravitational potential
field of the planet is now given by \Eq{eq:Phi_p}, with
\begin{equation}
M_p^* = M_p+M_{\rm CPD}\times f_b(s)\ ,
\label{eq:M_pert}
\end{equation}
so that inside $br_H$, the potential is basically unchanged. Outside, the
effective perturbing mass of the planet is $M_p+M_{\rm CPD}$.

Ideally, for $s<br_H$, one should use $M_{\rm H}(s/r_H)$ as defined by
\Eq{eq:mass} instead of $M_{\rm CPD}f_b(s)$. But this is
computationally more expensive as it requires calculating $M_{\rm
H}(s/r_H)$ for every cell of the grid within a distance $br_H$ to the
planet. Alternatively, one could also use a function other than $f_b$,
which would fit the curves displayed in \Figs{fig:Mcpd-s12_Jup} and
\ref{fig:Mcpd-s15_Jup} better.  For consistency, we have kept
$f_b$. This should have only a minor influence on the CPD, and even
less on the global simulation. However, it is important that a smooth
enough function is used\,: with a Heaviside-like function, $\Phi_p(s)$
could not be monotonic. In the present case, $f_{0.6}'(s)<4.2$, while
$M_{\rm CPD}/M_p\approx 0.01$, so that the change in $\Phi_p'(s)$ at
$s=0.6r_H$ does not exceed $4\%$.

With this prescription, we restarted our simulation from $t=350$
orbits, with the planet on a fixed circular orbit, and a high
resolution of $\delta r/r = \delta \theta = 2\times 10^{-3}$. The mass
of the CPD is added smoothly over 3 orbits to the planet mass to avoid
a discontinuity\,: in \Eq{eq:M_pert}, $M_{\rm CPD}$ is replaced by
$M_{\rm CPD}*g(t)$ with $g(t) = \sin^2 \left((t-350)\pi/6\right)$ for
$t<353$ and $g=1$ for $t\geq 353$.

\begin{centering}
\begin{figure}
\includegraphics[angle=0,width=\linewidth]{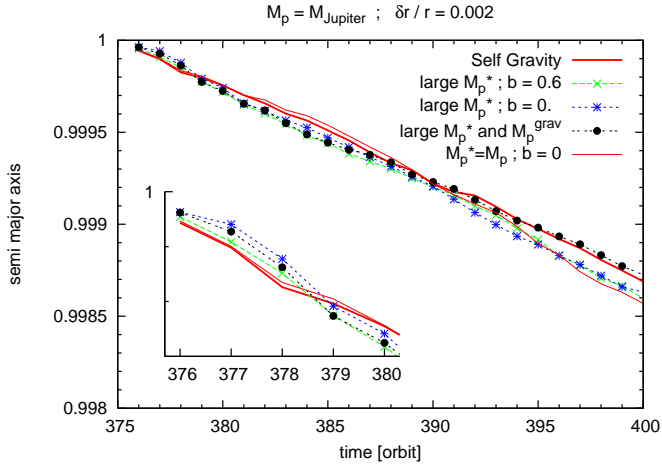}
\caption{Migration path of the Jupiter mass planet after its release
at $t=375$ orbits. The label ``large $M_p^*$'' denotes simulations for
which the planet perturbing mass is given by \Eq{eq:M_pert}. The label
``large ${M_p}^{\rm grav}$'' means that the gravitational mass of the
planet is $M_p+M_{\rm CPD}$. The bold curve is the reference case,
with full gas selfgravity (and $M_p^*=M_p$).}
\label{fig:migr-s15_Jup_Mcpd}
\end{figure}
\end{centering}

At $375$ orbits, the planet is released and free to migrate, like in
\Sect{sss:HR}. At this time, the CPD mass is $1.02\times 10^{-5}
M_*=1\%M_p$, like in \Sect{sss:HR}. The result in terms of migration
is displayed in \Fig{fig:migr-s15_Jup_Mcpd} as the curve labelled
``large $M_p^*$ ; b=0''. The thin red solid curve labelled
``$M_p^*=M_p\ ;\ b=0$'' is the most standard simulation, where nothing
is done (no full selfgravity, no exclusion, no modification of the
masses), also present in \Fig{fig:migr-s15_Jup}. Surprisingly, the
migration is slowed down in the first three orbits when the perturbing
mass of the planet is increased. The green curve with $\times$ symbols
corresponds to a case where $M_p^*$ is given by \Eq{eq:M_pert}, and in
addition $b=0.6$ in \Eq{eq:Force}. The black curve with big dots
corresponds to a simulation where the perturbing mass is given by
\Eq{eq:M_pert} and the gravitational mass of the planet is $M_p+M_{\rm
CPD}$. In the first five orbits, adding $M_{\rm CPD}$ to the
gravitational mass of the planet or excluding $60\%$ of the Hill
sphere both lead to a similar acceleration of the migration\,: the
green and black curves overlap at $t=380$, below the blue
curve. However, in the longer term (20 orbits), these two recipes lead
to different migration paths.

\subsection{Acceleration of the CPD}
\label{subsec:accCPD_Jup}

Last, we tested the recipe proposed by
\citet{Peplinski-etal-2008-I}. The perturbing mass of the planet is
given by \Eq{eq:M_pert}. The force felt by the planet is
$\vec{F_p}_{,\,0}$, given by \Eq{eq:Force}. The acceleration (or the
specific force) felt by the planet from the disc is then
$\vec{F_p}_{,\,0}/M_p$. This acceleration is then applied to the
material inside the Hill sphere, after multiplication by
$1-f_{0.5}(s)$, so the additional acceleration is smoothly vanishing
toward zero when $s>0.5\,r_H$. The material inside $0.5\,r_H$ feels
the same acceleration exactly as the planet and should naturally
follow the same orbit and the same migration path. This should set the
planet free of the ``ball and chain'' effect of the CPD.

\begin{centering}
\begin{figure}
\includegraphics[angle=0,width=\linewidth]{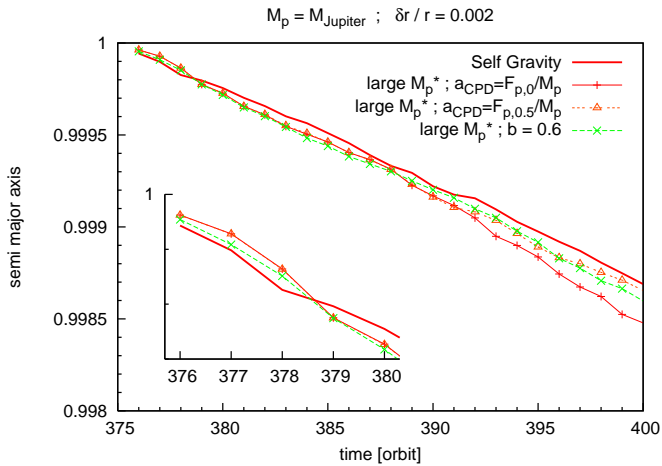}
\caption{Migration path of the Jupiter mass planet after its release
at $t=375$ orbits. The perturbing mass of the planet is
given by \Eq{eq:M_pert} in every case, except the full selfgravity case
(bold solid red line). The acceleration applied to the CPD varies.}
\label{fig:accCPD}
\end{figure}
\end{centering}

The resulting migration path is shown in \Fig{fig:accCPD} as the red
curve with $+$ symbols. The orange curve with open triangles is computed
with the same algorithm, except that the acceleration felt by the CPD
is the one exerted on the planet by the PPD only, as discussed in
\Sect{subsec:specific}\,: instead of $\vec{F_p}_{,\,0}/M_p$,
$\vec{F_p}_{,\,0.5}/M_p$ is applied to the CPD, but the planet still
feels $\vec{F_p}_{,\,0}$. During the first 15 orbits,
this has only a marginal influence on the migration and the two curves
overlap. But after $t=390$, the paths differ strongly. This proofs that our
reasoning in \Sect{subsec:specific} was not vain\,: it is important to
decide whether the acceleration applied to the CPD should include the
acceleration of the CPD on the planet or not. The path in the second
case is closer to the bold red reference full selfgravity case. We
recommend that the acceleration of the CPD on the planet should not be
applied to the CPD itself.

The green dashed curve with $\times$ symbols is taken from
\Fig{fig:migr-s15_Jup_Mcpd}\,: $M_p^*$ is given by \Eq{eq:M_pert} in
\Eq{eq:Phi_p}, and $b=0.6$ in \Eq{eq:Force}. The green and orange
curves are very close but not identical. The acceleration of the CPD
or its exclusion of the force exerted by the disc on the planet both
give the planet an inertial mass of $M_p$, but in a different way.

Accelerating the CPD is appealing, because this method gives the
planet an inertial mass of $M_p$, without having to exclude a part of
the disc. Moreover, it can be used with the addition of
$M_{\rm CPD}$ to the perturbing mass of the planet. However, it does
not give the same migration path as the gas full selfgravity either.

\subsection{Summary of the type~II migration case}

We observed the migration rate of a Jupiter mass planet, using various
numerical recipe to take the CPD into account. The mass of the CPD is
only $1\%$ of the planet mass here, but we find that the way the CPD
is considered can change the migration significantly. In particular at
lower resolution, excluding a part of the Hill sphere can lead to
variations of almost a factor two.

On the other hand, changing the perturbing or gravitational masses of
the planet or accelerating the inner half of the Hill sphere leads
to a moderate change in the migration speed. The resulting migration
path is compatible with the full selfgravity case. Unfortunately,
none of the methods leads to the same migration path as when the full
selfgravity of the gas is computed.

This suggests that the mass of the CPD was in this case too low
to strongly influence the migration, but that the outer layers of the
Hill sphere are able to exert a strong torque on the planet. Even
if the Hill sphere contains a small amount of gas, it should therefore
be considered with care. This is be discussed in more detail in
\Sect{sec:direct}.

\section{Type~III migration case}
\label{sec:numIII}

The second case that we studied is a Saturn mass planet with
$\Sigma_0(r)=10^{-3}r^{-1/2}\ M_*/{a_p}^2$. This density is one order
of magnitude higher than in previous case, so that type~III migration
is expected for this planet. Indeed, except for what concerns the
slope of the density profile, this corresponds to the case studied by
\citet{MassetPapaloizou2003}, who found type~III migration, and
\citet{DAngelo-etal-2005}, who claim that type~III migration does not
happen at high resolution. Another difference with these previous
works is that our equation of state is not locally isothermal, which
could significantly change the corotation torque
\citep{BaruteauMasset2008AD,KleyCrida2008}.

\subsection{Inertial mass problem}
\label{subsec:ExH-Sat}

The same experiment as in the Jupiter mass planet case was
performed. The setup was identical, with only the $M_p$ and $\Sigma_0$
changed.

\subsubsection{Middle resolution}

\begin{centering}
\begin{figure}
\includegraphics[angle=0,width=\linewidth]{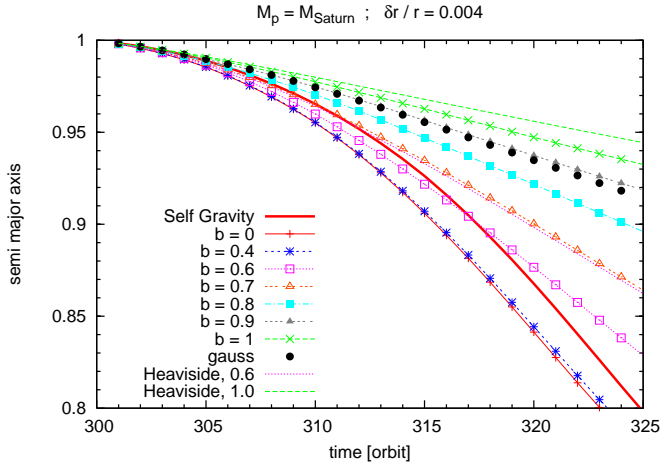}
\caption{Migration path of the Saturn mass planet after its release at
$t=300$ orbits. Bold curve\,: simulation where the gas selfgravity is
computed. Curves with symbols\,: simulations with exclusion of a part
of the Hill Sphere, using the filter $f_b$, for various values of $b$.
Curves without symbols\,: exclusion of a part of the Hill Sphere with
Heaviside filters. Big dots\,: exclusion of the Hill sphere with a
Gaussian filter.}
\label{fig:migr-s12_Sat}
\end{figure}
\end{centering}

For a restart at 300 orbits and resolution of $4\times 10^{-3} a_p =
r_H/11.4$, the results are displayed in \Fig{fig:migr-s12_Sat}. The
bold, red curve corresponds to the case where the full selfgravity of
the gas has been computed. In the other cases, only the axisymmetric
component of the selfgravity is computed. The curves with symbols
come from simulations with exclusion of a part of the Hill Sphere,
using the filter $f_b$, for various values of $b$ ($b=0$ means no
exclusion). For $b\leqslant 0.3$, the curves overlap and only the case
$b=0$ is drawn. Then, for increasing $b$, the migration speed
decreases.

The black dots correspond to a case where $f_b(s)$ has been replaced
in \Eq{eq:Force} by the Gaussian filter $g(s)$ given by \Eq{eq:g}. It
gives almost the same result than $f_{0.9}$ (grey blue short dashed
curve with full triangles).

The green dashed line without symbols corresponds to exclusion of the
Hill sphere with a Heaviside filter $h(s)=0$ if $s<r_H$, $h(s)=1$ if
$s>r_H$. It gives the most divergent result with the full selfgravity
case. The pink dotted line without symbols corresponds to exclusion of
the inner $60\%$ of the Hill sphere with a Heaviside filter
$h_{0.6}(s)=0$ if $s<0.6\,r_H$, $h_{0.6}(s)=1$ if $s>0.6\,r_H$. It
gives a result similar to $f_{0.7}$ (orange triple dashed line with
open triangles). Not only does the size of the excluded region matters
but also the shape of the filter.

First of all, we find that type~III runaway migration happens in that
case, in agreement with \citet{MassetPapaloizou2003}, even with full
selfgravity (bold red line). Admittedly however, we have not reached
the resolution of \citet{DAngelo-etal-2005} here (see next
section).

Second, when varying $b$, the migration rate changes dramatically, and
for $b\sim 1$, the runaway migration process seems to be broken. This
again shows that excluding a part of the Hill sphere or not can have
strong consequences on the migration. At $t=325$ orbits, the
semi-major axis of the planet is $a=0.77$ in the case $b=0$ and $0.93$
in the case $b=1$. The ratio of the semi major axis variations is
$3.33$.

Third, the observed behaviour is opposite the expected one if one
considers only the inertial mass effect\,: here the migration speed
decreases with $b$. The total mass of gas included in the Hill Sphere
at $t=300$ orbits is $2\times 10^{-5}\ M_* = M_p/16$. The impact on
the migration speed should thus not be more than $7\%$. This supports
the idea that the observed difference in the migration rate when
varying $b$ is not due to a change in the {\it inertial mass} of the
planet, but to the {\it direct torque} repartition inside the Hill
sphere. The mass and torque distribution inside the Hill sphere will
be discussed in \Sect{sec:direct}.

\subsubsection{Very high resolution}

\begin{centering}
\begin{figure}
\includegraphics[angle=0,width=\linewidth]{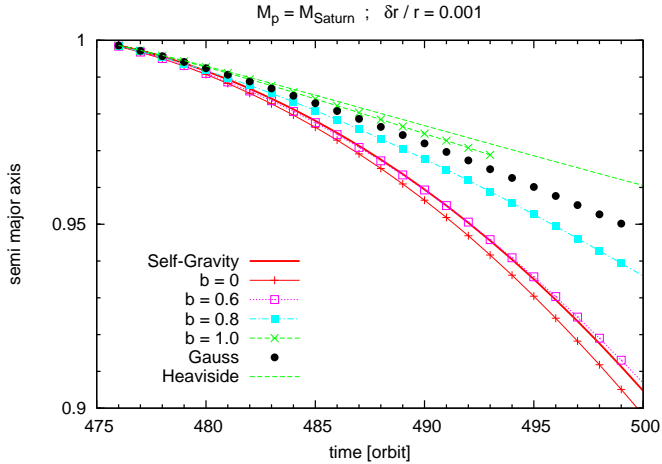}
\caption{Migration path of the Saturn-mass planet after its release at
$t=475$ orbits. Bold red curve\,: the full gas selfgravity is
computed. Other curves\,: only the axisymmetric part of the gas
selfgravity is computed. Thin solid red curve with $+$ symbols\,: the
whole disc is taken into account. The other curves come from simulations
with exclusion of a part of the Hill sphere, with the filter $f_b$ for
various values of $b$, for the Gaussian filter $g$, or for the Heaviside
filter $h$.}
\label{fig:migr-s19_Sat_ExH}
\end{figure}
\end{centering}

This simulation is run until $t=475$ orbits, with the planet on a fixed
circular orbit, and increasing resolution following
Table~\ref{tab:resol}. The axisymmetric component of the gas gravity
is always taken into account. In the end, a resolution of $10^{-3}$ is
reached, so that the length of a Hill radius is covered by $48.2$
cells. This is almost exactly the settings of the run 2D5Gb of
\citet{DAngelo-etal-2005}, for which convergence in resolution was
reached. The planet is then released and let free to migrate,
excluding a bigger or smaller part of the Hill sphere. The resulting
migration paths are displayed in \Fig{fig:migr-s19_Sat_ExH} for
various filters.

\begin{centering}
\begin{figure}
\includegraphics[angle=270,width=\linewidth]{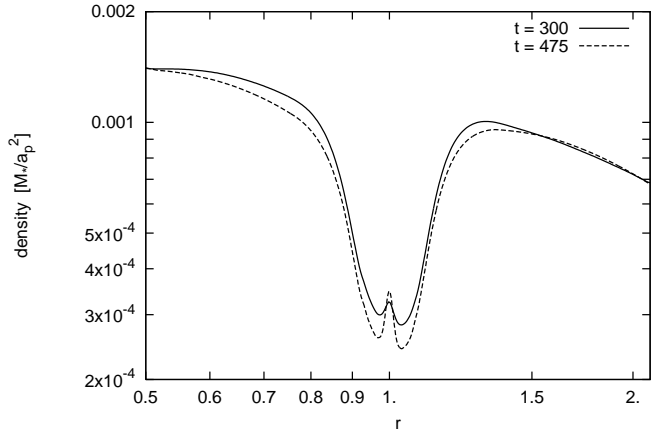}
\caption{Density profiles at $t=300$ orbits ($\delta r/r=\delta \theta
= 0.004$) and at $t=475$ orbits ($\delta r/r=\delta \theta = 0.001$),
with a Saturn mass planet on circular orbit at $r=1$.}
\label{fig:profiles_Sat}
\end{figure}
\end{centering}

The migration speed is much slower than at middle resolution
(\Fig{fig:migr-s12_Sat}), as already claimed by
\citet{DAngelo-etal-2005} (note the different $y$-axis scale), but the
migration rate exponentially increases, in a runaway typical of
type~III migration. In fact, the difference can arise from two
independent effects\,: (i) as already pointed out for the type~II
migration case, the disc density profile has evolved with time between
$t=300$ and $475$ orbits, independently of the resolution (see
\Fig{fig:profiles_Sat}). Therefore, the torque felt by the planet at
the release date differs. If it is smaller initially (as it appears to
be from \Fig{fig:profiles_Sat}), the torque remains smaller during the
runaway. (ii) The $e$-folding time of type~III migration scales as
$C\!M\!D/(C\!M\!D-M_p)$, where $C\!M\!D$ is the coorbital mass
deficit, and $M_p$ should be understood as the inertial mass of the
planet \citep[see][]{MassetPapaloizou2003}. Here, the total mass of
gas in the coorbital region in the unperturbed disc is about $6\times
10^{-4} M_*$, and the value of $C\!M\!D$ is of the order of
$M_p$. Therefore, a little increase in the inertial mass of the planet
(due to the increase in $M_{\rm CPD}$) results in a significant
increase in the $e$-folding time of the runaway migration. We think
that this explains most of the difference between
\Figs{fig:migr-s12_Sat} and \ref{fig:migr-s19_Sat_ExH}. The latter
effect may also be responsible for the vanishing of type~III migration
at high resolution in \citet{DAngelo-etal-2005} (Fig.~8) because the
mass inside the Hill sphere increases with resolution up to a few
$M_p$, according to their Fig.~9. This shows once again that the Hill
sphere should be considered with care. In the type~III migration
regime, convergence in the Hill sphere structure had better be
reached. Here, however, we are not looking for the real migration
rate, but for the influence of the gas in the Hill sphere on the
migration rate. This experiment is suitable for this purpose.

The major change observed does not stem for a strong modification of the
disc structure\,: the density profile is drawn in
\Fig{fig:profiles_Sat} at the two times where the planet is
released. The gap width and depth are similar. The CPD is almost twice
as massive at very high resolution than at moderate resolution,
though, reaching $1\%$ of the planet mass. This promotes the idea that
the material in the Hill sphere or the CPD plays a major role in the
type~III migration rate.

In the previous section, the results showed that a large part of the
torque responsible for the runaway migration originated in the Hill
sphere. Therefore, the lower migration rate observed in a
very-high-resolution case leads to the conclusion that the torque
exerted on the planet by the material inside the Hill sphere decreases
with resolution.

However, for the topic of this paper, we see that even at very high
resolution, excluding more than $60\%$ of the Hill sphere has a
significant influence on the migration speed. The larger the excluded
area, the more the migration diverges with respect to the full
selfgravity case. The Heaviside filter to exclude all what is beyond
$r_H$ of the planet appears to be the least appropriate.

\subsection{Perturbing and gravitational mass problem, acceleration of the CPD}

\subsubsection{Middle resolution}

With the same prescription as in \Sect{subsec:grav_Jup}, we
restarted our simulation from $t=250$ orbits, with the planet on a
fixed circular orbit and a resolution of $0.004$.  At this time,
$M_{\rm CPD}=M_{\rm H}(0.6)$ has reached $10^{-5}\,M_*=0.035\,M_p$. It
is added smoothly over 40 orbits to $M_p^*$, to avoid a discontinuity.

\begin{centering}
\begin{figure}
\includegraphics[angle=0,width=\linewidth]{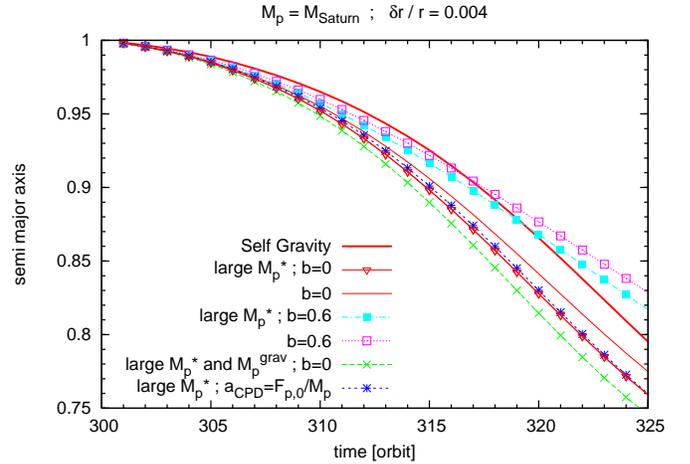}
\caption{Migration path of the Saturn mass planet after its release at
$t=300$ orbits, for different perturbing masses. The label ``large
$M_p^*$'' denotes simulations for which $M_p^*$ is given by
\Eq{eq:M_pert} since $t=290$. The bold curve labelled ``Self Gravity''
is computed with full gas selfgravity.}
\label{fig:m}
\end{figure}
\end{centering}

Then, we let the planet evolve freely under the influence of the disc
at time $t=300$ orbits. The migration path is shown in
Fig.~\ref{fig:m} as the red thin solid curve with triangles (labelled
``large $M_p^*$\,; b=0''), compared to the migration rate without
taking the CPD mass into account (thin red solid curve, labelled
``b=0''), and to the selfgravity case (bold solid curve). By excluding
in addition the inner $60\%$ of the Hill sphere ({\it i.e.}\ taking
$b=0.6$ in \Eq{eq:Force}), one gets the light blue, dot-dashed line
with full squares (labelled ``large $M_p^*$ ; b=0.6''). It can be
compared to the similar case, with $M_p^*=M_p$, which is the pink,
dotted line with open squares (labelled ``b=0.6''). The green dashed
curve with $\times$ symbols (labelled ``large $M_p^*$ and $M_p^{\rm
grav}$'') corresponds to the last case described in
\Sect{subsec:specific}\,: the perturbing mass is given by
\Eq{eq:M_pert}, and gravitational mass of the planet is $M_p+M_{\rm
CPD}$.

As expected, the migration is accelerated by the augmentation of the
perturbing mass of the planet. Adding $M_{\rm CPD}$ to the perturbing
mass of the planet leads to a rise in the migration rate of $4\%$ in
both cases $b=0$ and $b=0.6$. For the 15 first orbits, the curves
$a_p(t)$ remarkably overlap if one replaces $t$ by $300+(t-300)/1.04$
in the cases with $M_p^*=M_p$. Another $4\%$ increase in the migration
speed is observed when $M_{\rm CPD}$ is also added to the
gravitational mass of the planet. This is consistent with the mass of
the CPD.

The ``Self Gravity'' simulation was launched at 300 orbits
from the same disc as the other simulations here\,: before $t=300$
only the axisymmetric component of the selfgravity is computed, and
the disc is perturbed by a planetary potential given by \Eq{eq:Phi_p}
with $M_p^*$ given by \Eq{eq:M_pert}. The migration path is almost
indistinguishable from the one in the previous section, restarting from
the classical potential perturbed disc.

Unfortunately, as the full selfgravity case gives a slower migration
than the standard simulation, adding the mass of the CPD to the
perturbing and gravitational masses of the planet gives an even more
divergent result. This means that in this case, the selfgravity plays
another role as simply modifying the perturbing, gravitational, and
inertial masses of the planet.

%\paragraph{Acceleration of the CPD}

As in \Sect{subsec:accCPD_Jup}, the algorithm of
\citet{Peplinski-etal-2008-I} is also tested and compared to the other
ones. The migration obtained with it, with also $M_p^*$ given by
\Eq{eq:M_pert}, is shown in Fig.~\ref{fig:m}, as a blue curve with
stars, labelled ``large $M_p^*$ ; $a_{\rm CPD}$=$F_{p,0}/M_p$''. It
follows the path of the case with simply $M_p^*$ given by
\Eq{eq:M_pert}.

\subsubsection{Very high resolution}

\begin{centering}
\begin{figure}
\includegraphics[angle=0,width=\linewidth]{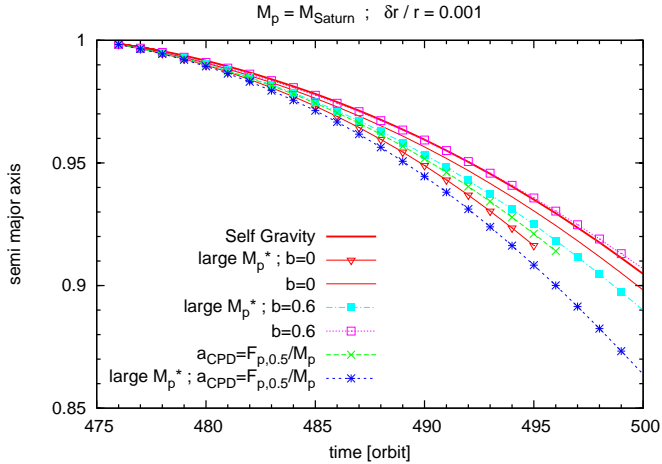}
\caption{Migration path of the Saturn mass planet after its release at
$t=300$ orbits. The label ``large $M_p^*$'' denotes simulations for
which $M_p^*$ is given by \Eq{eq:M_pert} since $t=290$. The bold curve
labelled ``Self Gravity'' is computed with full gas selfgravity.}
\label{fig:m19}
\end{figure}
\end{centering}

The same experiments are computed at very high resolution from $t=475$
orbits. At this time, the mass distribution in the Hill sphere of the
planet is shown in \Fig{fig:mass-torque-s19}, and $M_{\rm CPD} =
M_H(0.6) = 1.933\times 10^{-5}M_* = 0.068M_p$. For the cases with
$M_p^*=M_p+M_{\rm CPD}$, the simulation is restarted with this
perturbing mass at $t=450$ orbits, with the planet on a circular orbit
until 475 orbits, where the planet is released. As expected, the
addition of $M_{\rm CPD}$ to $M_p^*$ accelerates the migration. This
can be seen by comparing the thin solid red line to the thin solid red
line with open triangles in \Fig{fig:m19}. These two cases have $b=0$
but a different perturbing mass. During the 5 first orbits, the
migration paths overlap exactly if one replaces $t$ with
$475+(t-475)/1.068$ in the $M_p^*=M_p$ case. In the longer term,
however, the acceleration of the migration is closer to $10\%$ than
$6.8\%$.

With $b=0.6$, the influence of the perturbing mass can be seen by
comparing in \Fig{fig:m19} the pink dotted line with open squares with
the light blue dot-dashed line with full squares. The migration rates
compare exactly like in the $b=0$ case\,: $6.8\%$ difference during
the 5 first orbits, $10\%$ in the longer term.

An acceleration of $\vec{F_p}_{\,0.5}/M_p$ is then applied to the CPD
(green dashed curve with $\times$ symbols, labelled ``$a_{\rm
CPD}=\vec{F_p}_{\,0.5}/M_p$''). This leads to an overall acceleration
of the migration of about $7\%$ with respect to the standard, $b=0$
case. This could be expected, but is very different than excluding the
CPD from the computation of the force of the disc on the planet (pink
dotted curve with open squares). However, both methods aim at
delivering the planet from the inertia of the CPD. This shows that they
are not equivalent.

If in addition to the CPD acceleration, the perturbing mass is
$M_p^*=M_p+M_{\rm CPD}$ (blue dashed curve with stars), the migration
is accelerated again. The speed-up is exactly $6.8\%$ with respect
to the previous case (acceleration of the CPD with $M_p^*=M_p$) during
the first 5 orbits and about $8\%$ in the longer term. Last, in the case
where $M_p^*=M_p+M_{\rm CPD}$ and $b=0$, the acceleration of the CPD
leads to a speed-up in the migration of about $5\%$.

\subsection{Summary of the type~III migration case}

The outer layers of the Hill sphere ($s>0.6r_H$) play a big role in
the type~III migration regime. Excluding more than $60\%$ of the Hill
sphere leads to a decrease in the migration rate and even to a
suppression of the runaway phenomenon. In the full selfgravity
simulations, the runaway is present, even at very high resolution,
although the migration rate is decreased. Adding $M_{\rm CPD}$ to the
perturbing and/or gravitational masses of the planet clearly leads to
an acceleration of the migration at the expected rate. Accelerating
the CPD appears to have less of an effect at middle resolution, while
it plays a significant role at very high resolution.

At middle resolution, no recipe gives a perfect match with the
case where the full selfgravity of the gas is computed. We should
also conclude that the gas selfgravity plays a more complex role in
the type~III migration than simply accelerating the CPD or letting
the CPD perturb the PPD.

At very high resolution, excluding a part of the Hill sphere also has
a strong influence on the migration path. With $b=0.6$, the agreement
with the full selfgravity case is very good. However, this cannot be
explained by the analysis of \Sect{subsec:specific}, according to
which taking $b=0.6$ should accelerate the migration while it is
slowed down here. This agreement is therefore most likely a
coincidence. In any case, the type~III migration rate is strongly
reduced with increased resolution.

\section{Direct torque and Hill sphere structure}
\label{sec:direct}

\subsection{Direct torque from the CPD}

A short analysis of the direct torque from the CPD, defined as the
material bound to the planet, shows that it should be negligible. It
has been observed in numerical simulations with grid refinement
that two spiral arms form in the CPD, because of tidal effects from the
star. Thus, the CPD should not be considered as an axisymmetric
structure, but it looks centro-symmetric anyway. At lower resolution,
the spiral arms disappear\,; however, the axisymmetry of the structure
is not perfect. Only the perturbation from the axisymmerty can be
responsible from a torque on the planet.

It would be absurd that the CPD alone directly exerts a torque on the
planet and makes it migrate on the long term, because the CPD is
linked to the planet\,; this torque is an internal interaction in the
closed \{CPD+Planet\} system. This system cannot change its orbit by
itself. The only possible angular momentum transfer is between the
spin angular momentum of the CPD about the planet and the orbital
angular momentum of the \{CPD+Planet\} system about the star, similar
to the slow down of the Earth rotation together with increase of the
Moon-Earth distance. If one assumes for simplicity that the CPD is a
constant density disc of radius $r_{\rm CPD}$ in Keplerian rotation
about the planet, its spin angular momentum is
\begin{equation}
J_{\rm CPD} = \int_0^{r_{\rm CPD}} 2\pi s \Sigma \sqrt{GM_ps}\ {\rm
d}s = \frac{4}{5}M_{\rm CPD}\sqrt{GM_p}\sqrt{r_{\rm CPD}}\ .
\label{eq:Hcpd}
\end{equation}

The orbital angular momentum of the planet is
$J_p=M_p\sqrt{GM_*}\sqrt{a_p}$. The radius of the CPD is generally of
the order of $0.4r_H$. Thus $J_{\rm CPD}/J_p \approx 0.5 (M_{\rm
CPD}/M_p)q^{2/3}$. Even in the case of a CPD as massive as the planet
and of a 3 Jupiter mass planet, the ratio would be inferior to
$1\%$. In the cases studied in Sects.~\ref{sec:numII}
and~\ref{sec:numIII}, one finds $J_{\rm CPD}/J_p \approx 5\times
10^{-4}$. Thus, the spin angular momentum of the CPD is negligible
with respect to the orbital angular momentum of the planet, and
consequently the tidal effects of the star (or the PPD) on the latter
can only have a negligible effect on the planetary migration.

\subsection{Mass and torque distribution inside the Hill sphere}

\begin{centering}
\begin{figure}
\includegraphics[width=\linewidth,angle=0]{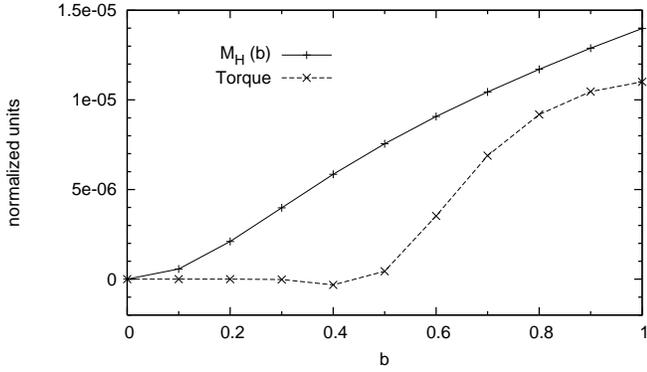}
\caption{Mass $M_{\rm H}(b)$ (solid line, \Eq{eq:mass}) and Torque
(dashed line) distribution in the Hill sphere of the Saturn mass
planet at time $t=300$ orbits.}
\label{fig:mass-torque}
\end{figure}
\end{centering}

We consider the situation after 300 orbits in the case of a Saturn
mass planet in a massive disc for the middle
resolution. Figure~\ref{fig:mass-torque} displays the mass inside the
Hill sphere $M_{\rm H}(b)$. It increases significantly
with $b$ even for $b>0.4$, because the gap opened by the planet is not
empty (see \Fig{fig:profiles_Sat})\,; the background density is
responsible for the increase in $M_{\rm H}$ with $b$. However, the
azimuthally averaged profiles shown in \Fig{fig:profiles_Sat} show
that a CPD is present.

The torque exerted on the planet by the region within $s<br_H$, is
also plotted in \Fig{fig:mass-torque} as a function of $b$. More
precisely, it is the torque of the force computed with \Eq{eq:Force},
with $(1-f_b(s))$ instead of $f_b(s)$\,; or equivalently the torque of
the force\,: $\vec{F_p}_{,\,0}-\vec{F_p}_{,\,b}$. It appears that for
a planet on a fixed circular orbit, the region inside $r_H/2$ exerts a
negligible torque, while the region between $r_H/2$ and $r_H$ has a
non negligible effect on the planet\,: the {\it direct torque} is not
small.

\begin{centering}
\begin{figure}
\includegraphics[width=\linewidth,angle=0]{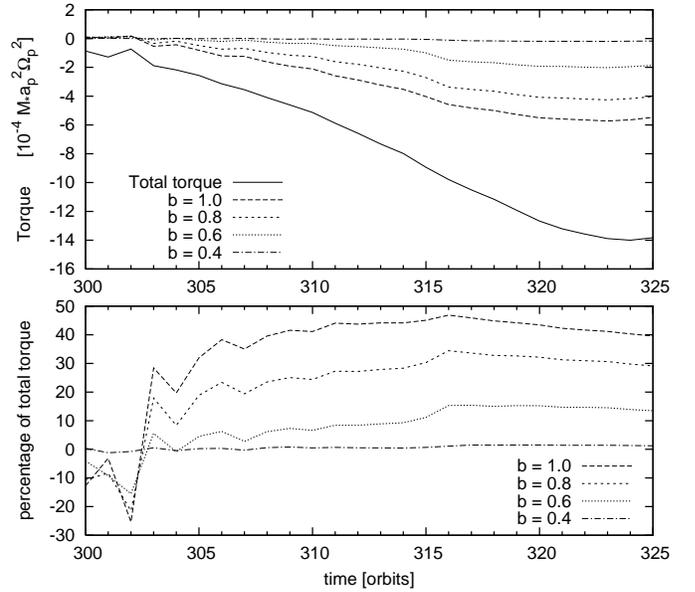}
\caption{Torque exerted on the planet as a function of time, during
its migration, by all the disc (bottom, solid line), the Hill sphere
(dashed line $b=1.0$), the inner $80\%$ of the Hill sphere ($b=0.8$),
$60\%$ ($b=0.6$), and $40\%$ ($b=0.4$). Bottom panel\,: the same
torques as a percentage of the total torque.}
\label{fig:torques}
\end{figure}
\end{centering}

Figure~\ref{fig:torques} shows the torque exerted on the planet by
various regions of the disc, as a function of time, during the
type~III migration between 300 and 325 orbits. The measures are done
in the simulation with full gas selfgravity computed. The total
torque, exerted by all the disc, is increasingly negative, which is
characteristic of runaway inward migration (solid line). The other
curves are the torques due to spheres centred on the planet inside the
Hill sphere, like in \Fig{fig:mass-torque}. In the bottom panel, the
torques due to fractions of the Hill sphere are plotted in terms of
percentage of the total torque. The torque exerted by the material
inside the Hill sphere is positive before the planet is released, but
then turns negative. The inner $40\%$ of the Hill sphere plays a
negligible role for a migrating planet too. But the entire Hill sphere
is responsible for about $40\%$ of the total torque. This confirms
that the region between $s=r_H/2$ and $s=r_H$ plays a major role in
the process of type~III migration, as \Fig{fig:migr-s12_Sat} suggests.

\begin{centering}
\begin{figure}
\includegraphics[width=\linewidth,angle=0]{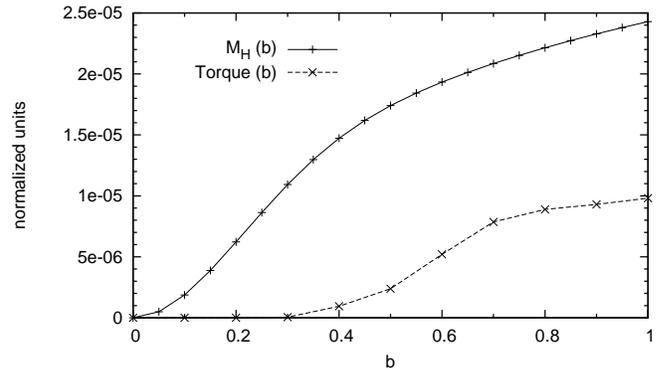}
\caption{Mass $M_{\rm H}(b)$ (solid line, \Eq{eq:mass}) and torque
(dashed line) distribution in the Hill sphere of the Saturn mass
planet at time $t=475$ orbits.}
\label{fig:mass-torque-s19}
\end{figure}
\end{centering}

\begin{centering}
\begin{figure}
\includegraphics[width=\linewidth,angle=0]{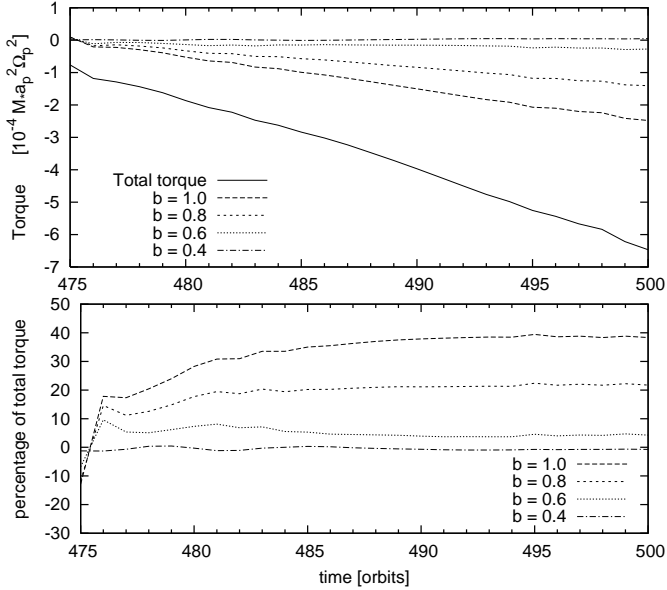}
\caption{Torque exerted on the planet as a function of time, during
its migration, by all the disc (bottom, solid line), the Hill sphere
(dashed line $b=1.0$), the inner $80\%$ of the Hill sphere ($b=0.8$),
$60\%$ ($b=0.6$), and $40\%$ ($b=0.4$). Bottom panel\,: the same
torques as a percentage of the total torque.}
\label{fig:torques-s19}
\end{figure}
\end{centering}

Figures~\ref{fig:mass-torque-s19} and \ref{fig:torques-s19} are the
same as \Figs{fig:mass-torque} and \ref{fig:torques}, but from time
$475$ orbits on, with the very high resolution. The torques as a
function of time in \Fig{fig:torques-s19} come from the simulation
with only the axisymmetric component of the selfgravity taken into
account, and no exclusion of any part of the Hill sphere. The two
figures look very similar to the previous ones, in particular the
total torque is smaller but is also increasingly negative. The other
previous remarks also apply at very high resolution.

\subsection{Size of the circum-planetary disc}
\label{sec:size}

The calculation results and the analysis of the mass and torque
distributions inside the Hill sphere suggest that the region within
$0.5\,r_H$ of the planet contains most of the mass and yet it exerts a
negligible torque on the planet, whereas the region between $0.5\,r_H$
and $r_H$ has a negligible mass but it exerts a large torque on the
planet. Such a large torque cannot be permanently exerted by material
bound to the planet, otherwise this material would not migrate at the
same rate as the planet's, and it would ultimately escape from the
planet's potential well. This suggests that the CPD is located inside
$0.5\,r_H$, which can be checked with a streamline analysis, as shown
below. Alternatively, one can determine the shape and the size of the
CPD with the following energetic approach. We denote by $E_{\rm tot}$
the total specific energy of the fluid elements perturbed by the
planet. It is the sum of the potential, kinetic, and thermal perturbed
energies by unit mass of the fluid elements\,:
\begin{equation}
E_{\rm tot} = -GM_p/s' + |\vec{v}-\vec{v_p}|^2/2  +  (T-T_0)/(\gamma-1) \ ,
\label{eq:E_tot}
\end{equation}
where $\vec{v}-\vec{v_p}$ is the velocity with respect to the planet,
and $T-T_0$ is the perturbed temperature with respect to the initial
state. The minimum energy that the fluid elements must have to leave
the planet's potential well with zero velocity with respect to the
planet defines the escape energy, denoted by $E_{\rm esc}$. It reads
\begin{equation}
E_{\rm esc} = -GM_p/d_{\rm stag}' +  (T-T_0)_{d_{\rm stag}}/(\gamma-1) \ ,
\label{eq:E_esc}
\end{equation}
with $d_{\rm stag}' = \sqrt{d^2_{\rm stag} + \epsilon^2}$ the
smoothed separation between the planet and one of the two hyperbolic
stagnation points (where the velocity of the flow with respect to the
planet cancels out). Fluid elements such that $E_{\rm tot} < E_{\rm
esc}$ should be bound to the planet, and thus be part of the CPD,
whereas those having $E_{\rm tot} \geq E_{\rm esc}$ should orbit the
central star. Said differently, the shape and the size of the CPD
should correspond to the locations in the disc where $E_{\rm tot} =
E_{\rm esc}$.

%------------------------
\begin{centering}
  \begin{figure}
%    \resizebox{\hsize}{!}
%    {
      \includegraphics[width=\linewidth,angle=0]{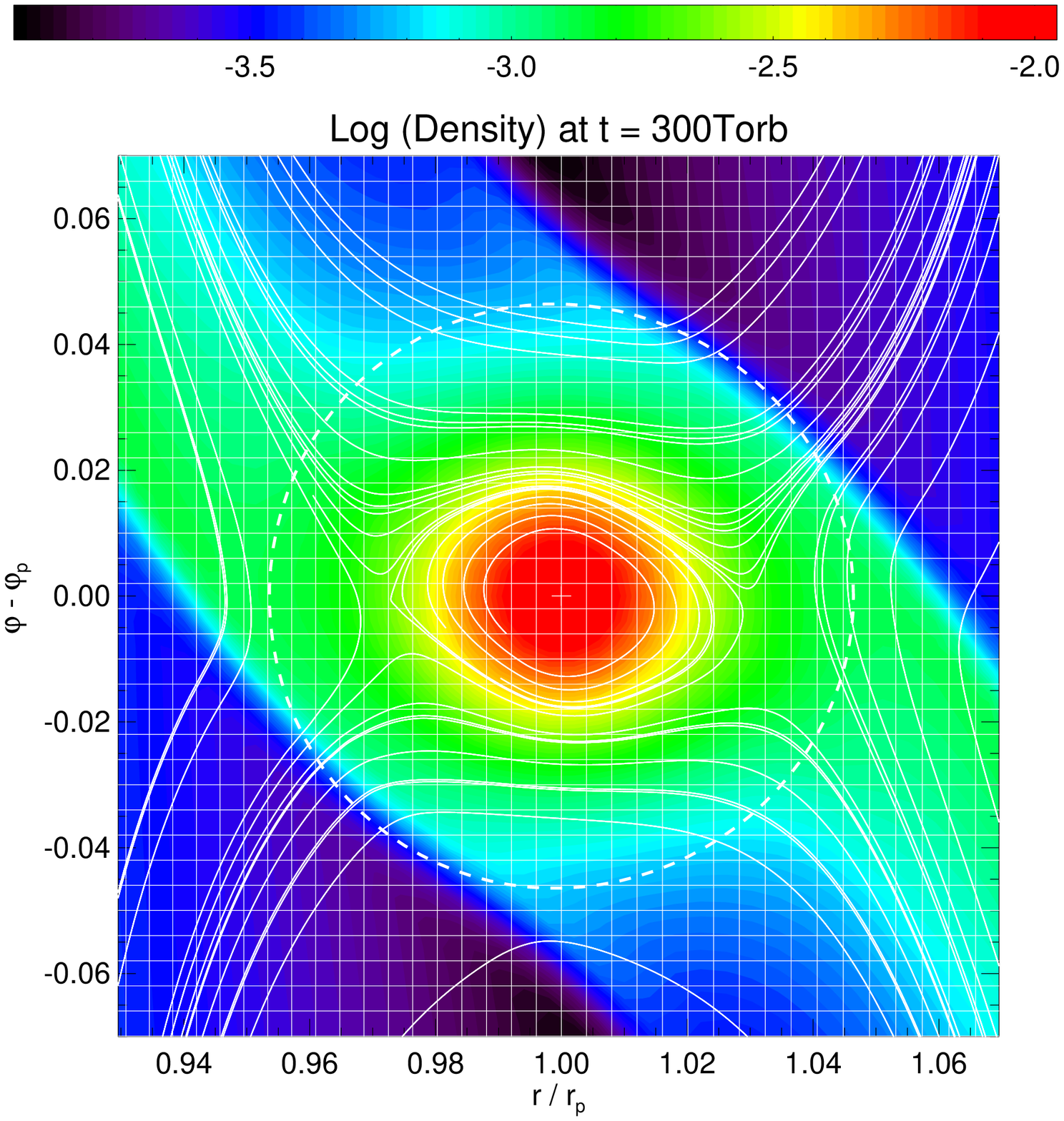}
      \includegraphics[width=\linewidth,angle=0]{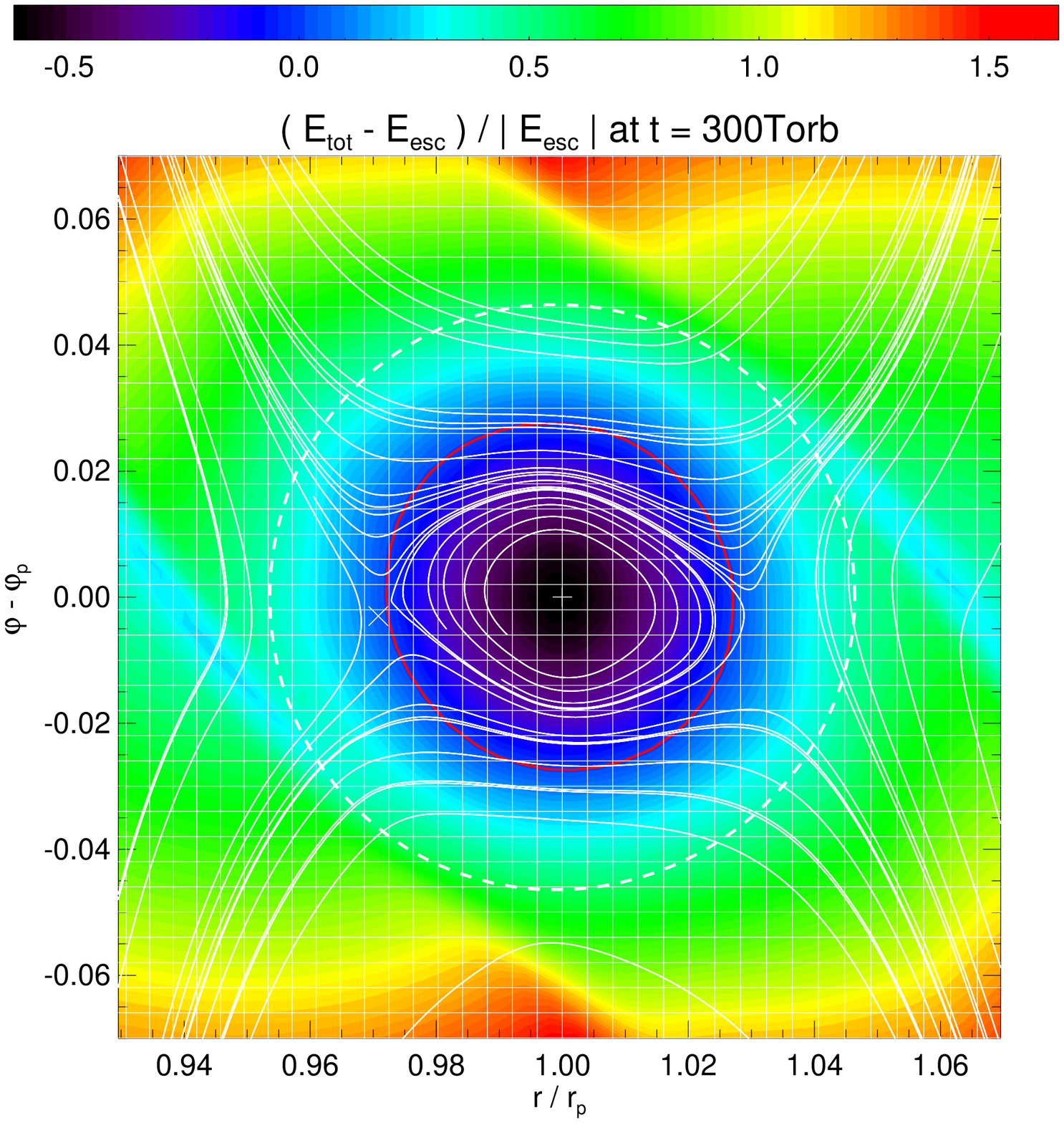}
%    } 
    \caption{Top\,: surface density contour at $300$ orbits for the
    middle-resolution run with the Saturn-mass planet. The planet and
    its Hill sphere are respectively represented by the $+$ symbol and
    the dashed circle. The grid mesh is superimposed, as well as
    streamlines in the planet's frame. Bottom\,: same as the top panel,
    except the quantity $(E_{\rm tot}-E_{\rm esc})/|E_{\rm esc}|$ is
    now displayed. The red solid curve shows the contour $E_{\rm
    tot}=E_{\rm esc}$, and the $\times$ symbol shows the location of
    the stagnation point used to evaluate the escape energy $E_{\rm
    esc}$.}
    \label{fig:maps_SATMR}
  \end{figure}
\end{centering}
%------------------------

We first compare both approaches to determine the CPD shape with the
middle-resolution run for the Saturn mass planet. In the top panel of
\Fig{fig:maps_SATMR}, we depict the gas surface density at $300$
orbits, just before the planet release. The latter is located at
$r=r_p$ and $\varphi = \varphi_p$, and its position is highlighted
with a $+$ symbol. The background vertical and horizontal lines show
the grid mesh. The dashed circle represents the planet's Hill
sphere. Streamlines (solid curves) are also superimposed to appreciate
the shape and the size of the CPD. The streamline analysis confirms
that most of the mass inside the planet's Hill sphere is confined to
the CPD, and it shows that the latter has an elliptical shape, with a
semi-major axis approximately twice as large as the semi-minor
axis. The semi-major axis is $\sim 0.6\, r_H$. This justifies that the
material inside $\sim 0.5\, r_H$ exerts a small torque on the planet
compared to the material located between $0.5\,r_H$ and $r_H$.

In the bottom panel of \Fig{fig:maps_SATMR}, we display the quantity
$(E_{\rm tot}-E_{\rm esc})/|E_{\rm esc}|$ at the same time. The same
streamlines as in the top panel are depicted. The red solid line shows
the contour $E_{\rm tot}=E_{\rm esc}$, the quantity $E_{\rm esc}$
being calculated with Eq.~(\ref{eq:E_esc}). The location of the
stagnation point, depicted with a $\times$ symbol in this panel (at $r
\approx 0.97r_p$, $\varphi \approx \varphi_p$), is inferred from the
streamline analysis. The contour $E_{\rm tot}=E_{\rm esc}$ can be
approximated as a circle of radius $0.6\, r_H$. Its location agrees
with that of the CPD determined with the streamline analysis.

\begin{centering}
  \begin{figure}
%    \resizebox{\hsize}{!}
%    {
      \includegraphics[width=\linewidth,angle=0]{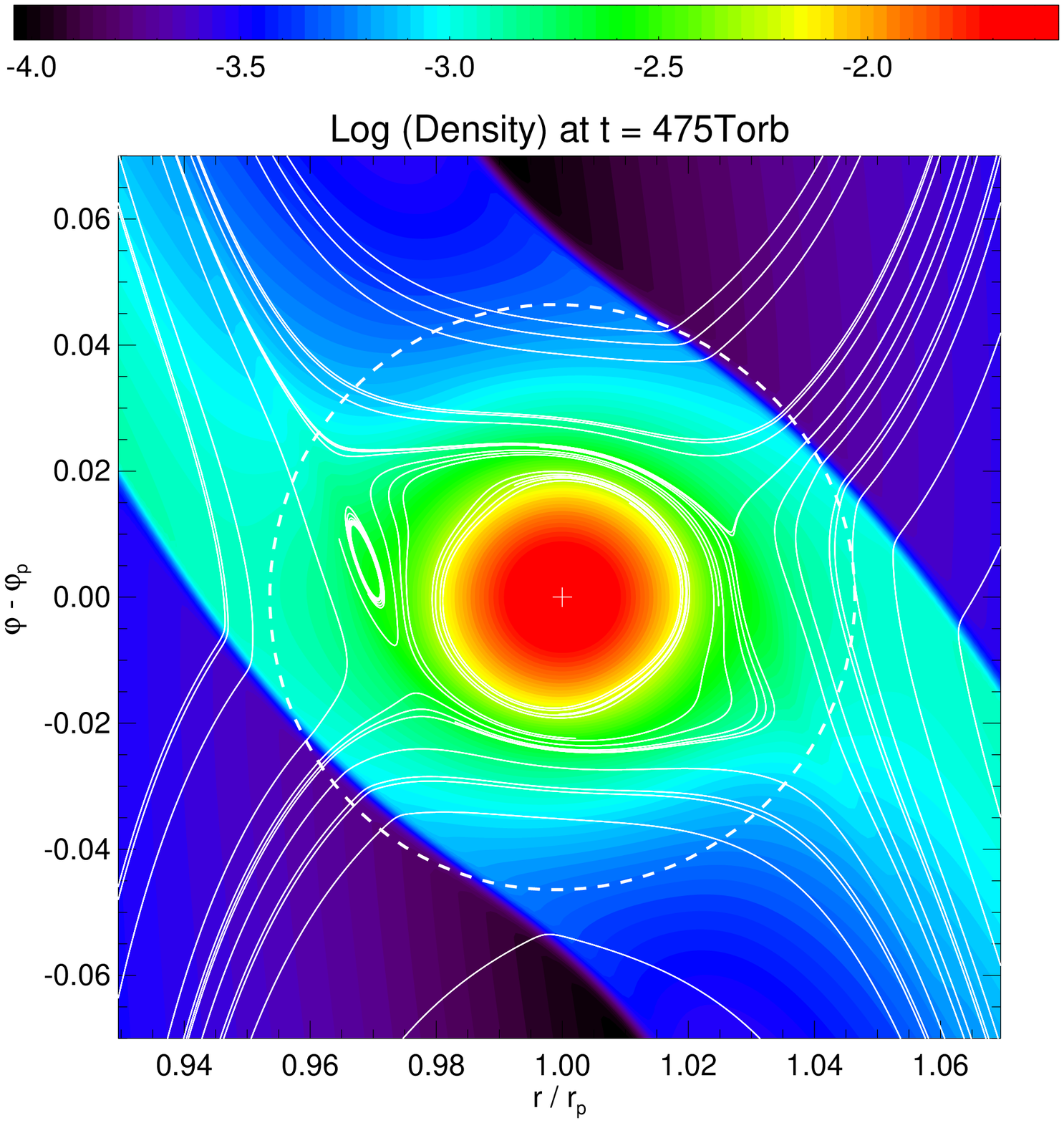}
      \includegraphics[width=\linewidth,angle=0]{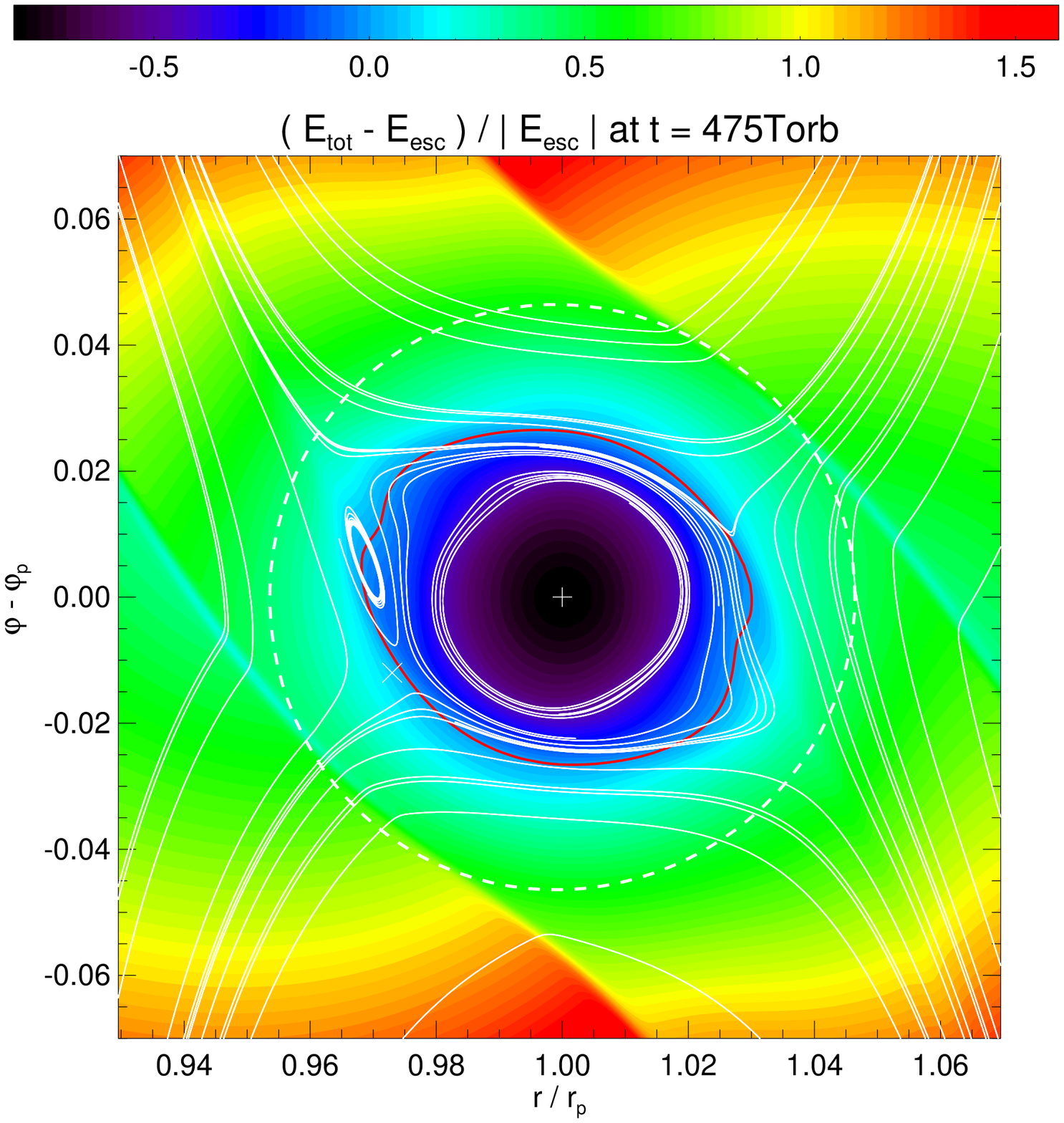}
%    } 
    \caption{Same as \Fig{fig:maps_SATMR}, but for the
    very-high-resolution run, just before the release of the planet at
    475 orbits. To improve legibility, the grid mesh is not
    displayed.}
    \label{fig:maps_SATVHR}
  \end{figure}
\end{centering}

Figure~\ref{fig:maps_SATVHR} is the same as \Fig{fig:maps_SATMR}, but for the
very high-resolution run just before the release of the Saturn mass
planet at $475$ orbits. A close comparison between both figures
reveals significant differences in the flow topology near the
planet. In particular, the CPD is now very close to a circle of radius
$0.5\,r_H$, and its shape and its size are in very good agreement with
those of the contour $E_{\rm tot}=E_{\rm esc}$. The flow also tends to
show more complexity at higher resolution, judging from the presence
of a single vortex slightly inside the planet's orbit.

%------------------------
%------------------------

The CPD is slightly larger than in previous locally isothermal
studies by \citet{DAngelo-etal-2005}. This probably comes from the
structure inside the Hill sphere changeing when one
changes the equation of state, as the pressure support is
modified. With the energy equation, the collapse of the CPD is limited by
the heating due to adiabatic compression, which may give a wider but
less massive CPD than in the locally isothermal case. Also, the smoothing
length $\epsilon$ influences the mass and size of the CPD\,: the
smaller $\epsilon$, the deeper the potential well of the planet, and
then the larger the CPD. The choice of the equation of state, opacity,
and smoothing length certainly influences the size and mass of the
CPD. The analysis performed in this section concerns only the specific
cases that we studied in this paper, in order to compare the mass and
torque repartition in the Hill sphere with the migration rates
observed.

\section{Conclusion}
\label{sec:conclu}

In this study, we have performed experiments to analyse the role of
the circum-planetary disc and the gas in the Hill sphere in the
numerical simulations of planetary migration of giant planets. The way
the material inside the Hill sphere is taken into account has a strong
influence on the outcome of simulations in which the gas selfgravity
is not computed. Whether or not a part of the Hill sphere is excluded
in the computation of the force felt by the planet, the migration rate
can vary by a factor 2 in type~II migration, and even $3.33$ in case
of type~III migration. This numerical issue is thus
critical. Therefore, we would like to encourage authors to specify how
they deal with the Hill sphere of giant planets in their simulations,
so that their experiments can be reproducible by others.

In \Sect{subsec:specific}, we analysed the problem and proposed
several solutions. To deliver the migrating planet from the ball and
chain effect of the CPD, it is possible to exclude a part of the Hill
sphere from the computation of the force exerted by the disc on the
planet. It is also possible to artificially give the CPD the
acceleration felt by the planet. Both methods are not equivalent. In
order to take the CPD into account in the perturbation of the PPD, the
mass of the CPD can be added to the perturbing mass of the planet. In
the second case studied here, this leads to an acceleration of the
migration proportional to the mass of the CPD. To help the planet to
pull the CPD, the gravitational mass of the planet can be increased by
$M_{\rm CPD}$, which also accelerates the migration as expected. These
four tricks can be combined.

We would like to point out again that the CPD is not the entire Hill
sphere. As demonstrated in \Sect{sec:size}, its size is only about a
half of the Hill radius only. As a consequence, excluding all the Hill
sphere of the planet is inappropriate\,; in fact, our simulations show
that this method gives the most divergent results with respect to our
reference case (full gas selfgravity computed). Unfortunately, no
method matches the full selfgravity case. The role of the selfgravity
is more complex than changing the inertial, perturbing, and
gravitational masses of the planet.

The present study was conducted for a specific equation of state
and for a specific value of the softening length. Under these
circumstances, it seems appropriate to remove about half
of the Hill sphere in the computation of the force felt by the planet
from the disc, but not more than $0.6r_H$. Authors who consider
different prescriptions should undertake a prior analysis of the CPD
size to infer how they should scale this fraction in their
case. Alternatively, accelerating the material within $\sim\!0.5\,r_H$
of the planet also provides satisfactory results, but the acceleration
imposed on the CPD should be the one provided to the planet by the
disc outside $0.5r_H$. The addition of $M_{\rm CPD}$ to the perturbing
mass of the planet also seems to be a good idea and leads to an
acceleration of the migration. Of course, the most reliable solution
is to compute the full gas selfgravity, with the highest resolution
possible. In any case, one should be aware of the strong influence of
the CPD on planetary migration.

\begin{acknowledgements}

A. Crida acknowledges the support through the German Research
Foundation (DFG) grant KL 650/7. The high-resolution computations were
performed on the {\tt hpc-bw} and {\tt hpc-uni} clusters of the
Rechenzentrum of the University of T\"ubingen.

\end{acknowledgements}

\bibliographystyle{aa}
\bibliography{crida}

\end{document}